\documentclass[preprint,12pt,authoryear]{elsarticle}

\usepackage{amssymb}
\usepackage{amsmath}
\usepackage{booktabs}
\usepackage{pdflscape}
\usepackage{threeparttable}
\usepackage{graphicx}
\usepackage{float}
\usepackage{adjustbox}
\usepackage{xcolor}
\usepackage{afterpage}
\usepackage[a4paper,left=1.25in, right=1.25in, top=1.4in, bottom=1.4in ]{geometry}
\usepackage{hyperref}
\journal{International Journal of Forecasting}

\begin{document}
\vspace{-5em}

\begin{frontmatter}

\title{Distributional Forecasting of EU Asylum Applications with Dynamic Multivariate Count Models}


\author[inst1]{Gregor Zens\corref{cor1}}
\ead{zens@iiasa.ac.at}
\author[inst2]{Jakub Bijak}
\ead{jakub.bijak@reuben.ox.ac.uk}

\cortext[cor1]{Corresponding author. Address: International Institute for Applied Systems Analysis (IIASA), Schlossplatz 1, 2361 Laxenburg, Austria.}
\affiliation[inst1]{organization={International Institute for Applied Systems Analysis},
            city={Laxenburg},
            country={Austria}}
\affiliation[inst2]{organization={University of Oxford},
            city={Oxford},
            country={United Kingdom}}

\begin{abstract}
We propose a Bayesian framework for joint distributional forecasting of monthly asylum applications across the EU-27. The model decomposes latent application intensities into country-specific random walks and common factors, with idiosyncratic and shared shocks allowed to exhibit heavy tails or stochastic volatility. Using Eurostat data from 2008 to 2026, we evaluate predictive distributions in a rolling out-of-sample exercise, scoring overall distributional accuracy and upper-tail risk. Three findings emerge. First, the preferred specification varies across countries, scoring rules, and horizons, underscoring the need to align models with policy-specific loss functions. Second, joint EU-27 models improve on country-by-country benchmarks, with the largest gains in the upper tail, where preparedness costs are most relevant. Third, random-walk log-intensities provide a useful short-run description of national asylum-application dynamics, especially when combined with flexible innovation dynamics. We conclude by discussing implications for national and EU-level agencies involved in asylum forecasting and preparedness planning.
\end{abstract}





\end{frontmatter}

\section{Introduction}

Asylum-related migration is central to political discourse in Europe and a recurring driver of policy debate at both national and European Union (EU) levels. Because sudden surges can strain reception capacity, administrative processing, and local service provision, preparedness and anticipatory planning are critical. This need has been explicitly recognised at the level of the EU with the adoption of the \textit{Migration Preparedness and Crisis Blueprint} \citep{blueprint}, following the high-intensity migration events of the mid-2010s. 

To support these policy objectives, a growing forecasting literature has emerged aimed at enhancing early warning and contingency planning. This strand of work focuses on a variety of approaches -- from classical time series methodologies like ARIMA models to change-point detection and machine learning-augmented early warning methods (\citealp{carammia2022forecasting}; \citealp{napierala2022}; \citealp{bosco2024machine}; \citealp{boss2025forecasting}; \citealp{wycoff2025forecasting}). In practice, several European countries and international agencies now produce regular short-term forecasts or early-warning assessments to support preparedness planning. An overview of this literature, methods, and good national and international practices has recently been compiled by the \cite{oecd2026}.

Despite this progress, at least three challenges remain for short-run asylum forecasting in Europe. First, many existing approaches focus on point prediction and evaluate models based on central-tendency error metrics, such as the root mean square error (RMSE). However, \textit{preparedness} is inherently distributional in this context, as the costs of forecast errors are strongly asymmetric: under-estimating inflows can create humanitarian bottlenecks and political crises, whereas over-estimating inflows typically results in temporary financial costs from maintaining idle beds or administrative capacity. From a statistical decision-theoretic perspective, such asymmetric costs make upper quantiles more appropriate prediction targets than measures of central tendency.\footnote{Under asymmetric linear loss, the Bayes-optimal point forecast is a predictive quantile, with the quantile level determined by the relative costs of under- and over-prediction.}

Second, recent forecasting approaches increasingly rely on complex machine-learning pipelines and rich covariate sets. While such methods can improve predictive performance, they often depend on inputs that are difficult to measure consistently across countries, costly to update in real time, and challenging to operationalize in practice. Their black-box nature can also complicate the communication of model structure and forecast results to policymakers. These concerns are amplified by the limited length of asylum application time series, which contain only around 200 monthly observations per country. In such settings, parsimonious probabilistic models remain attractive from both a statistical and operational perspective.

Third, asylum flows to different destination countries are frequently modelled as separate national processes, even though they are interconnected and correlated. Events in origin countries, such as the outbreak of armed conflict, can increase applications across several destinations simultaneously, while changes in asylum policy in one country may redirect flows toward others, making destinations partly substitutable \citep{dehaasetal2019,czaikaetal2025}. Country-specific models therefore risk neglecting cross-country dependence that may be especially relevant during periods of shared shocks.

We address these shortcomings by introducing a parsimonious model for joint distributional forecasting of asylum applications, with a latent structure informed by migration theory. To move beyond point prediction, the framework produces full predictive distributions for future application counts. To reduce reliance on difficult-to-update external inputs, it models the dynamics of asylum applications directly, without requiring covariates that must themselves be forecast. To account for cross-country interdependence, movements in latent application intensities are decomposed into common shocks and idiosyncratic country components, with both components allowed to exhibit heavy tails or stochastic volatility. This structure reflects the empirical instability of asylum flows and their dependence across destination countries, while remaining sufficiently transparent for policy-oriented forecasting applications.

Our contribution is both empirical and methodological. Empirically, we provide -- to the best of our knowledge -- the first systematic distributional forecast evaluation for EU asylum applications, using scoring rules that map naturally into distinct preparedness objectives, including full-distribution accuracy and upper-tail risk. Methodologically, we contribute to the literature on forecasting multivariate count time series (e.g., \citealp{aktekin2018sequential}; \citealp{berry2020bayesian}) by adapting dynamic factor stochastic-volatility ideas to this setting. Our model can be viewed as a Poisson state-space analogue of sparse factor stochastic-volatility models \citep{kastner2019sparse}, in which nonstationary latent log-intensities evolve with a time-varying covariance structure. For computation, we develop a custom Markov chain Monte Carlo (MCMC) estimation strategy that accommodates the Poisson observation equation, missing observations, rolling forecast evaluation, and multi-step predictive simulation.

We evaluate the model using monthly European asylum application data from January 2008 to February 2026. In a rolling pseudo out-of-sample forecasting exercise, we assess distributional predictive performance using scoring rules tailored to distinct policy objectives, capturing both full-distribution accuracy and upper-tail risk.

Three main findings emerge. First, consistent with the complex nature of migration processes, we find no universally optimal model for different policy objectives. Instead, the preferred specification depends on the destination country, the forecast horizon, and the evaluation metric -- and thus on the policymaker's operational loss function. For instance, a model that performs well in terms of overall distributional accuracy is not necessarily the best for extreme-capacity planning. These findings caution against relying on centralized, one-size-fits-all strategies for anticipatory action and motivate making the relevant loss functions as explicit as possible in both model evaluation and policy use.

Second, joint EU-27 models frequently improve on simpler independent country-by-country benchmarks, underscoring the merits of a joint treatment of migration and asylum processes across multiple ``competing'' destinations. This indicates that even models for national capacity planning benefit from considering the EU as an interconnected system and allowing forecasts to borrow strength, especially in times of global shocks.

Finally, a Poisson state-space framework where latent log-intensities evolve as random walks with flexible innovation dynamics provides a useful specification for short-run European asylum forecasting. This aligns with \citet{bijak2010}, who emphasizes the non-stationarity and limited long-term predictability of migration processes. This class of models generates well-calibrated predictive distributions, particularly in the upper tail, without depending on extensive external covariates that are difficult to update and operationalize in real time. This suggests that over short horizons, recent application levels hold substantial predictive information, provided the random-walk structure is combined with innovation dynamics flexible enough to accommodate volatility shifts, structural breaks, and unusually large inflows. Related evidence from the analysis of high-frequency irregular maritime migration in Europe points in the same direction \citep{zens2026dynamic}. 

The remainder of the article is structured as follows. Section~\ref{sec:framework} introduces the statistical framework. Section~\ref{sec:data} describes the data. Section~\ref{sec:prediction} presents the forecast experiment and empirical results. Section~\ref{sec:discussion} concludes. Additional implementation details are provided in \ref{app:mcmc}, while \ref{app:multistep-tables} reports additional figures and multi-step country-level results.

\section{Statistical Framework}
\label{sec:framework}
\subsection{Theoretical Considerations \& Empirical Specification}

Let \(z_{it}\in\mathbb{R}\) denote the latent log intensity of the number of asylum applications in the destination country \(i\) in time period \(t\). In accordance with classical migration theories, we assume that \(z_{it}\) is mainly driven by a factor \(a_{it}\) describing the attractiveness of a country -- summarizing its policy stance, economic welfare, existing migrant networks, etc. -- as well as a term \(g_{it}\) that reflects the impact of EU-wide and origin-region developments such as overall attractiveness of the EU as a destination or conflict intensity in origin regions. We assume a simple additive decomposition of the form
\[
z_{it}=g_{it}+a_{it}.
\]

Conceptually, this model specification draws inspiration from the \textit{generation -- distribution} tradition \citep{willekensbaydar1986,willekens1994,debeer2011}. In this framework, overall migration flows, including asylum, can be decomposed into \textit{generation} effects, driving the overall migration volumes, and \textit{distribution} effects, governing the allocation of migrants into individual destinations.\footnote{Alternative approaches for decomposing migration flows into global, origin and destination factors exist; notably, they include \textit{multiplicative components} approaches \citep[e.g.][]{stillwell1986,raymeretal2006,wisniowski2025}. However, the existing models are typically used for coarse temporal data and do not explicitly account for the volatility dynamics in the series.}

We model the \textit{distribution} effects describing country-specific attractiveness \(a_{it}\) of destination $i$ as random walks, reflecting that these are potentially nonstationary trending patterns that are hard to predict and evolve slowly over time $t$,

\[
a_{it}=a_{i,t-1}+u_{it}.
\]

The \textit{generation} effect of global shocks \(g_{it}\) is modeled based on the idea that we can decompose it into a set of \(Q\) \textit{factors}, \(f_{qt}\), corresponding to global migration driver environments \citep[see][]{czaikareinprecht}, where each factor describes a development (or a set of related developments) that is shared across destination countries \(i\). In line with the considerations in \citet{zens2026dynamic}, we assume these factors are also evolving according to random walks
\[
f_{qt}=f_{q,t-1}+v_{qt},\qquad q=1,\dots,Q.
\]
For parsimony, we assume that each country-specific intensity \(z_{it}\) has a constant loading for each of these latent series, denoted by \(\lambda_{iq}\), describing how important factor \(q\) is for asylum application intensity in country \(i\). Overall, this setup results in a latent factor structure of the form
\[
z_{it}=a_{it}+\sum_{q=1}^Q \lambda_{iq} f_{qt}.
\]

It remains to specify the law driving the innovations \(v_{qt}\) and \(u_{it}\), respectively describing  the period changes in global factors driving asylum into Europe (e.g., changes in conflict intensity in origin regions, EU-wide policy shifts, or global disruptions such as COVID-19) -- the generation effects -- and the period changes in country-specific attractiveness (e.g., changes in asylum laws, economic developments, or shifts in diaspora networks) -- the distribution effects. For now, we remain agnostic about these and simply assume that they can be represented conditionally as zero-mean Gaussian innovations with time-varying variances $\psi_{it}$ and $\omega_{qt}$:
$$
u_{it}\mid \psi_{it} \sim \mathcal{N}(0,\psi_{it}),
\qquad
v_{qt}\mid \omega_{qt} \sim \mathcal{N}(0,\omega_{qt}),
$$
so that each series \(i\) has its own time-varying idiosyncratic variance process \(\psi_{it}\), while each factor \(q\) has its own time-varying variance process \(\omega_{qt}\). This flexible structure accommodates complex real-world dynamics, such as long periods of limited change followed by short episodes of rapid adjustment. For example, on the country level, legislative turnover may be clustered, or economic crises may temporarily increase volatility in destination attractiveness before calmer periods resume. For global factors, conflict intensity may be relatively stable for some time and then change abruptly when a war front shifts; similarly, pandemic-related developments may induce a short period of unusually large common shocks.

This setup implies the following process for differences in log asylum intensity $\Delta z_{it}=z_{it}-z_{i,t-1}$:
$$
\Delta z_{it}
=
(a_{it}-a_{i,t-1})
+\sum_{q=1}^Q \lambda_{iq}\bigl(f_{qt}-f_{q,t-1}\bigr)
=
u_{it}+\sum_{q=1}^Q \lambda_{iq} v_{qt}.
$$

Stacking $\Delta z_{it}$ across $K$ destination countries yields $\Delta \boldsymbol{z}_t=(\Delta z_{1t},\dots,\Delta z_{Kt})'$. Let $\boldsymbol{\Lambda}=(\lambda_{iq})\in\mathbb{R}^{K\times Q}$, $\boldsymbol{u}_t=(u_{1t},\dots,u_{Kt})'$, and $\boldsymbol{v}_t=(v_{1t},\dots,v_{Qt})'$. Then
$$
\Delta \boldsymbol{z}_t = \boldsymbol{u}_t + \boldsymbol{\Lambda}\boldsymbol{v}_t,
\qquad
\boldsymbol{u}_t\mid \boldsymbol{\Psi}_t \sim \mathcal{N}(0,\boldsymbol{\Psi}_t),
\qquad
\boldsymbol{v}_t\mid \boldsymbol{\Omega}_t \sim \mathcal{N}(0,\boldsymbol{\Omega}_t),
$$
with diagonal covariance matrices
$$
\boldsymbol{\Psi}_t=\mathrm{diag}(\psi_{1t},\dots,\psi_{Kt}),
\qquad
\boldsymbol{\Omega}_t=\mathrm{diag}(\omega_{1t},\dots,\omega_{Qt}).
$$

Because the innovations are conditionally Gaussian, integrating out the latent factor innovations $\boldsymbol{v}_t$ yields a highly tractable reduced-form joint distribution for the differences $\Delta \boldsymbol{z}_t$:
$$
\Delta \boldsymbol{z}_t \mid (\boldsymbol{\Psi}_t,\boldsymbol{\Omega}_t)
\sim
\mathcal{N}\!\bigl(0,\boldsymbol{\Sigma}_t\bigr),
\qquad
\boldsymbol{\Sigma}_t=\boldsymbol{\Psi}_t+\boldsymbol{\Lambda}\boldsymbol{\Omega}_t\boldsymbol{\Lambda}'.
$$

Hence, changes in asylum application intensity are modeled as zero-mean but potentially correlated across EU destination countries, with a correlation (and covariance) structure that varies over time through \(\boldsymbol{\Sigma}_t\). Periods in which a strongly common shock dominates -- such as a pandemic affecting all destinations simultaneously -- can generate high positive cross-country correlation, whereas other periods may be characterized by mostly idiosyncratic movements (low correlation), by correlation concentrated within subsets of countries (e.g., clusters of destinations with similar policy regimes or geographic proximity), or by negative correlation, if for instance asylum seekers respond to restrictive policies by shifting routes and changing destination countries  \citep{dehaasetal2019,czaikaetal2025}, all depending on which latent factors are volatile at time \(t\) and how strongly countries load on them.

In reality, observed asylum applications $y_{it}$ are counts. We connect the latent intensity process to the observed application counts through a Poisson observation equation,
$$
y_{it}\sim \mathrm{Poisson}\bigl(\exp(z_{it})\bigr),
$$
yielding a flexible framework for modeling nonstationary multivariate count processes. In this sense, the model can be understood as a multivariate random-walk Poisson model with a dynamic covariance structure. The random-walk component captures the persistent evolution of asylum application rates, while the volatility and factor components allow the predictive distribution to adapt to periods of instability and shared shocks.

\subsection{Error Distributions and Volatility Dynamics}

To close the model, we must specify the distributional forms and the dynamic evolution of the idiosyncratic variances, $\psi_{it}$, and the factor variances, $\omega_{qt}$. Ultimately, we view the choice of the error distribution and volatility dynamics -- alongside the determination of the number of latent factors, $Q$ -- as an empirical problem. Because our primary objective is to produce reliable predictions, we refrain from imposing a single structural assumption \textit{a priori}. Instead, we rely on an out-of-sample forecasting exercise described in more detail in Section~\ref{sec:prediction}. We use out-of-sample predictive evaluations to select the score-preferred specification and factor dimension $Q$ within the candidate set. We consider three distinct parametric choices -- as well as their combinations -- for the innovation variances:

\paragraph{Gaussian (Homoskedastic)} 
The simplest specification assumes that the underlying shocks are strictly Gaussian with constant variances over time. Under this regime, $\psi_{it} = \sigma_{i}^2$ and $\omega_{qt} = \sigma_{q}^2$ for all $t$. This formulation yields a static covariance matrix $\boldsymbol{\Sigma}$, which is suitable for modeling systems where structural volatility is stable and the data generating process is not substantially heavy-tailed.

\paragraph{Student-$t$ (Heavy Tails)} 
To accommodate data with more extreme innovations without permanently inflating the variance of calmer periods, we consider a setting where innovations follow a Student-$t$ distribution \citep[as per][]{bijak2010}. 
For this, the idiosyncratic variance is parameterized as $\psi_{it} = \sigma_i^2 / \chi_{it}$, where $\chi_{it} \sim \mathcal{G}(\nu_i/2, \nu_i/2)$ is an independent gamma-distributed mixing variable, and $\nu_i$ represents the degrees of freedom. A similar scale mixture can be applied to the factor variances. This conditionally Gaussian representation retains the tractability of the factor structure while rendering the marginal distributions heavy-tailed.

\paragraph{Stochastic Volatility (SV)} 
To capture persistent volatility clustering -- where long periods of limited change are followed by episodes of rapid adjustment -- we consider modeling the variances as time-varying processes, as proposed in irregular migration contexts by \citet{zens2026dynamic}. Following standard formulations, the log-variances are modeled as persistent AR(1) processes \citep{kastner2016dealing}. Let $h_{it}=\log \psi_{it}$ and $h_{qt}=\log \omega_{qt}$. We assume
$$
h_{it} = \mu_i + \phi_i\,(h_{i,t-1}-\mu_i) + \sigma_{h,i}\,\xi_{it}, \qquad i=1,\dots,K,\qquad(countries)
$$
$$
h_{qt} = \mu_q + \phi_q\,(h_{q,t-1}-\mu_q) + \sigma_{h,q}\,\xi_{qt}, \qquad q=1,\dots,Q, \qquad(drivers)
$$
with $\xi_{it},\xi_{qt}\sim \mathcal{N}(0,1)$ being mutually independent. This explicitly allows both the overarching global trends and the country-specific baselines to experience smoothly varying periods of high and low volatility.

\subsection{Prior Setup}

We complete the Bayesian specification by assigning priors to the factor loadings and the variance components. For the factor loadings $\lambda_{iq}$, we adopt factor-specific horseshoe shrinkage priors \citep{carvalho2010horseshoe} to allow for sparse covariance patterns while retaining the flexibility for a subset of countries to load strongly on particular factors. For each factor $q$ and country $i$, we specify the horseshoe based on the standard auxiliary variable representation:
$$
\lambda_{iq}\mid \tau_q,\lambda_{iq}^{hs} \sim \mathcal{N}\!\bigl(0,\ \tau_q \lambda_{iq}^{hs}\bigr),
\qquad
\lambda_{iq}^{hs}\sim \mathcal{C}^+(0,1),
\qquad
\tau_q\sim \mathcal{C}^+(0,1).
$$
This prior tightly shrinks irrelevant factor loadings toward zero, encouraging a parsimonious factor structure, while the heavy tails of the half-Cauchy distributions permit large loadings when supported by the data.

For the idiosyncratic and factor innovations, our prior assignments depend on the chosen error specification:

\paragraph{Gaussian (Homoskedastic)} 
Under the homoskedastic Gaussian specification, the country-specific idiosyncratic variances are assigned a weakly informative Inverse-Gamma prior, $\sigma_i^2 \sim \mathcal{IG}(2.5, 1.5)$. To identify the scale of the unobserved factors relative to the loadings, the homoskedastic factor variances are fixed to unity, such that $\sigma_q^2 = 1$ for all $q$.

\paragraph{Student-$t$} 
When assuming Student-$t$ distributed errors, the degrees of freedom parameters $\nu_i$ and $\nu_q$ are assigned a shifted exponential prior, specifically $(\nu - 3) \sim \mathrm{Exp}(1/6)$. This restricts the degrees of freedom to be strictly greater than 3 -- ensuring finite second and third moments -- and yields a prior mean of 9. The baseline scales mirror the Gaussian specification: $\sigma_i^2 \sim \mathcal{IG}(2.5, 1.5)$ for the idiosyncratic errors, while the baseline factor scales are fixed to 1 to maintain scale identification.

\paragraph{Stochastic Volatility} 
In the stochastic volatility specification, the log-variance AR(1) parameters $(\mu, \phi, \sigma_h)$ follow the standard priors established in \citet{kastner2016dealing}. Specifically, we impose a $\mathrm{Beta}(5,1.5)$ prior on the transformed persistence parameter $(\phi+1)/2$ and a $\mathcal{G}(0.5,0.5)$ prior on the volatility of volatility $\sigma_h^2$. For scale identification of the factor system under stochastic volatility, we restrict the log-volatility means $\mu$ of the factors to be zero for $q=1,\dots,Q$. This effectively centers the baseline factor variances around one and pins down the scale of the latent factors, improving statistical identification and numerical stability of the estimation algorithm.

\subsection{Posterior Simulation}

Posterior inference is carried out using a custom Metropolis-within-Gibbs
sampler. For a training sample of length \(T\) and forecast horizon \(H\), the
latent log-intensity path is augmented with both an initial state and \(H\)
future states,
\[
\boldsymbol{z}_{i,0:S}
=
(z_{i0},z_{i1},\ldots,z_{iT},z_{i,T+1},\ldots,z_{i,T+H})',
\qquad
S=T+H .
\]
The Poisson likelihood is attached only to observed in-sample counts. The
initial state, missing observations, and appended future states are treated as
latent quantities, so multi-step prediction is embedded directly in the MCMC
state vector. At each retained draw, count forecasts are obtained from
\[
y^{\ast}_{i,T+h}
\sim
\mathrm{Poisson}\{\exp(z_{i,T+h})\},
\qquad h=1,\ldots,H .
\]
Full conditional distributions and implementation details are provided in
\ref{app:mcmc}.

\section{Data}
\label{sec:data}

We utilize official monthly asylum application counts from Eurostat, restricting our analysis to the current EU-27 member states. The sample period spans January 2008 through February 2026, yielding $T=218$ monthly observations across $K=27$ destination countries. The Croatian series starts in January 2013, the year in which Croatia joined the EU.\footnote{Under our proposed Bayesian estimation framework, this unbalanced panel structure poses no difficulties; the unobserved pre-2013 Croatian data points are treated as latent quantities and integrated out within the MCMC sampler.}

\begin{figure}[!htpb]
\centering
\includegraphics[width=\linewidth]{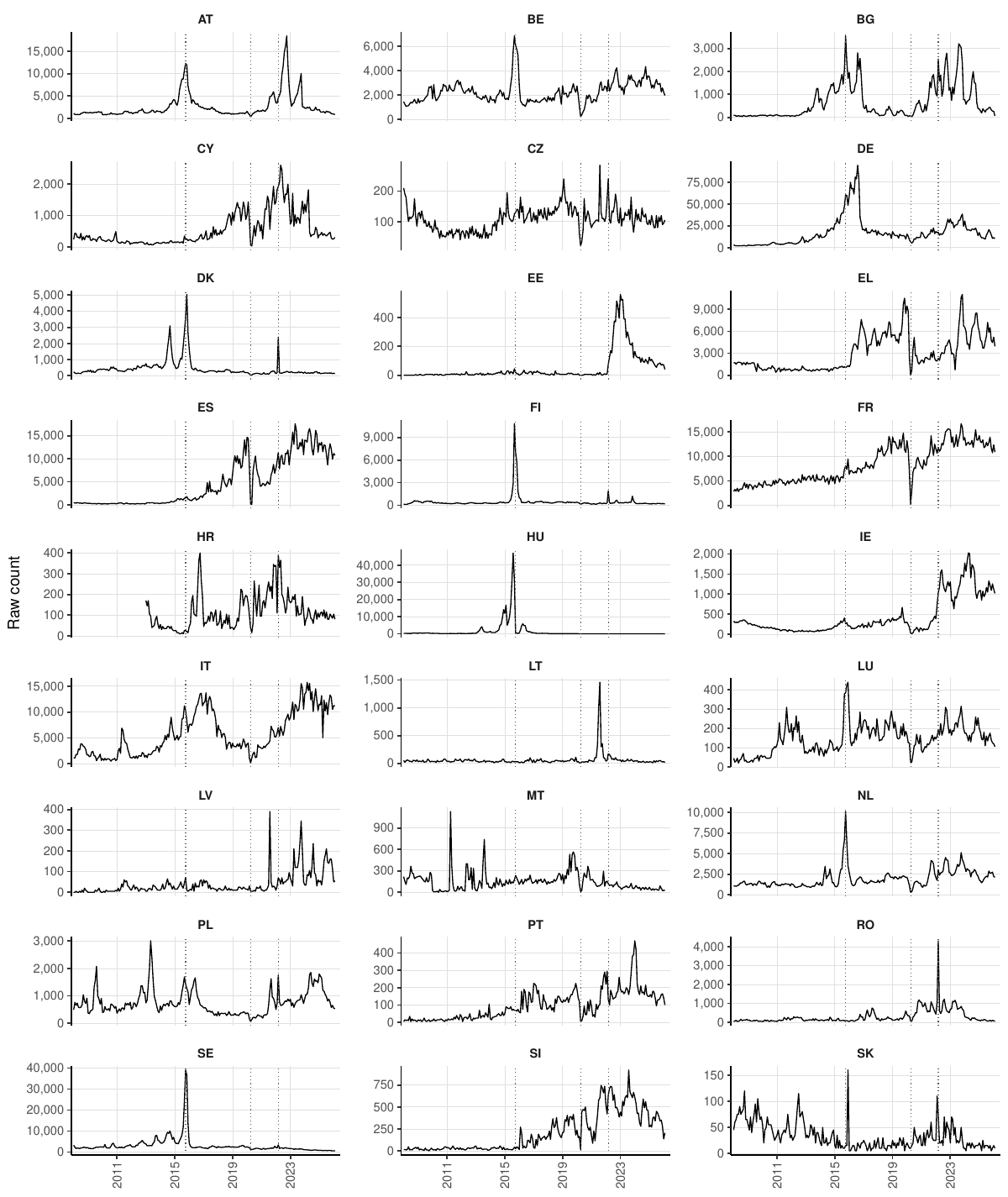}
\caption{Time series of raw monthly asylum application counts across the EU-27 (January 2008--February 2026). Vertical dashed lines denote major structural breaks and exogenous shocks: the peak of the Middle Eastern refugee surge (October 2015), the onset of COVID-19 mobility restrictions (April 2020), and the beginning of Ukrainian refugee inflows following the Russian invasion (March 2022). Data source: Eurostat.}
\label{fig:data-series}
\end{figure}

Figure~\ref{fig:data-series} plots the raw monthly asylum application counts. The corresponding log-transformed series, on the scale used by the latent
state process, are shown in Figure~\ref{fig:data-series-log} in the supplementary material. Visual inspection reveals two prominent empirical features characteristic of EU asylum flow data. First, the series are highly persistent, exhibiting pronounced level shifts and complex trends across multiple destinations. Second, the data demonstrate substantial, time-varying volatility. Extended periods of relative stability are punctuated by sharp structural breaks and rapid surges, such as the 2015--2016 influx of Middle Eastern refugees into specific EU states. In other periods, shocks manifest systematically across the panel; the onset of the COVID-19 pandemic in early 2020, for instance, induced a severe, synchronized contraction in asylum applications due to widespread mobility restrictions.

These stylized facts -- high persistence, volatility clustering, and dynamic cross-sectional dependence -- further motivate the statistical framework outlined in Section~\ref{sec:framework}, which explicitly accommodates a non-stationary log-intensity with time-varying innovation variances and a dynamic covariance structure. Country-level summary statistics are provided in Supplementary Table~\ref{tab:sumstats}.

\section{Results}
\label{sec:prediction}

\subsection{Forecast experiment}

We assess predictive performance using a rolling pseudo out-of-sample design. At each forecast origin, the models are estimated on a training window of 113
monthly observations and are used to generate predictive distributions up to six
months ahead for each EU-27 destination country. The estimation window is then
advanced by one month. This procedure yields 100 forecast evaluations per
country at each horizon. The one-month-ahead targets span June 2017 to
September 2025, while the six-month-ahead targets extend to February 2026.

The model comparison varies the distributional specification of both the
idiosyncratic innovations, \(u_{it}\), and the common-factor innovations,
\(v_{qt}\). For each component, we consider homoskedastic
Gaussian errors, Student-\(t\) errors, and Gaussian stochastic volatility as candidate specifications.  The case \(Q=0\) corresponds to a model without latent common factors. For
\(Q=1,\ldots,6\), we estimate all combinations of idiosyncratic and factor
innovation specifications. This results in 57 candidate models. For each model
and forecast origin, posterior inference is based on 60{,}000 retained MCMC
draws after an initial burn-in period of 6{,}000 iterations.

\subsection{Scoring rules and loss functions}

Let \(h\) denote the forecast horizon. For country \(i\), forecast origin \(t\),
and horizon \(h\), let \(y_{i,t+h}\) be the realized number of asylum
applications, let \(\widehat F^{(h)}_{i,t}\) denote the corresponding
\(h\)-step-ahead predictive distribution, and let
\(\widehat q^{(h)}_{i,t}(\tau)\) denote its predictive \(\tau\)-quantile. We
evaluate predictive performance using four complementary scoring rules (\citealp{gneiting2007probabilistic}; \citealp{czado2009predictive}) and loss functions. The
first two assess the full predictive distribution, while the latter two focus on
upper-tail risk, which is most relevant for preparedness planning.

First, we consider log predictive scores, defined as
\[
\mathrm{LPS}^{(h)}_{i,t}
=
\log \widehat p^{(h)}_{i,t}(y_{i,t+h}),
\]

where \(\widehat p^{(h)}_{i,t}(y_{i,t+h})\) is the predictive probability assigned to the realized count under \(\widehat F^{(h)}_{i,t}\). Larger values indicate better predictive performance.

Second, we evaluate continuous ranked probability scores,
\[
\mathrm{CRPS}^{(h)}_{i,t}
=
\mathbb{E}_{\widehat F^{(h)}_{i,t}}
\left|Y^{\ast}_{i,t+h}-y_{i,t+h}\right|
-
\frac{1}{2}
\mathbb{E}_{\widehat F^{(h)}_{i,t}}
\left|Y^{\ast}_{i,t+h}-\widetilde Y^{\ast}_{i,t+h}\right|,
\]
where \(Y^{\ast}_{i,t+h}\) and \(\widetilde Y^{\ast}_{i,t+h}\) are independent
draws from \(\widehat F^{(h)}_{i,t}\). This representation is valid for
discrete predictive distributions and is estimated directly from the posterior
predictive draws. Equivalently, because the outcome is a nonnegative count, the
CRPS can be written as
\[
\mathrm{CRPS}^{(h)}_{i,t}
=
\sum_{k=0}^{\infty}
\left\{
\widehat F^{(h)}_{i,t}(k)
-
\mathbf{1}(y_{i,t+h}\leq k)
\right\}^{2}.
\]

Third, we evaluate upper-tail risk using quantile-based losses. For a generic quantile level \(\tau\), the pinball loss -- also known as the check or asymmetric linear loss -- is defined as
\[
\rho_{\tau}
\{y_{i,t+h}-\widehat q^{(h)}_{i,t}(\tau)\}
=
\begin{cases}
\tau\{y_{i,t+h}-\widehat q^{(h)}_{i,t}(\tau)\},
& y_{i,t+h}\geq \widehat q^{(h)}_{i,t}(\tau),\\[0.4em]
(1-\tau)\{\widehat q^{(h)}_{i,t}(\tau)-y_{i,t+h}\},
& y_{i,t+h}<\widehat q^{(h)}_{i,t}(\tau).
\end{cases}
\]
For \(\tau>0.5\), this asymmetric loss penalizes underprediction more strongly than overprediction, making it suitable for evaluating forecasts used in high-capacity preparedness planning. We compute an upper-tail quantile score by integrating pinball losses over high quantile levels:
\[
\mathrm{UTQS}^{(h)}_{i,t}
=
\int_{0.90}^{0.999}
\rho_{\tau}
\{y_{i,t+h}-\widehat q^{(h)}_{i,t}(\tau)\}
\,d\tau.
\]
This score is a quantile-based upper-tail analogue of the CRPS and is closely related to the upper-tail component in the quantile representation of the CRPS. In the empirical application, the upper-tail score is approximated numerically over the grid
\[
\mathcal{T}_{U}
=
\{0.90,0.91,\ldots,0.98,0.985,0.990,0.995,0.996,0.997,0.998,0.999\}.
\]
Let the ordered grid points be
\[
0.90=\tau_1<\tau_2<\cdots<\tau_J=0.999, \qquad J=16.
\]
The upper-tail quantile score is then computed using the trapezoidal rule:
\[
\mathrm{UTQS}^{(h)}_{i,t}
=
\sum_{j=1}^{J-1}
\frac{1}{2}
\left[
\rho_{\tau_j}
\{y_{i,t+h}-\widehat q^{(h)}_{i,t}(\tau_j)\}
+
\rho_{\tau_{j+1}}
\{y_{i,t+h}-\widehat q^{(h)}_{i,t}(\tau_{j+1})\}
\right]
(\tau_{j+1}-\tau_j).
\]

Finally, to evaluate high-inflow risk in a scenario where policymakers target a specific tail quantile, we evaluate the pinball loss at the 95th percentile,
\[
\rho_{0.95}
\{y_{i,t+h}-\widehat q^{(h)}_{i,t}(0.95)\}.
\]
This score focuses on a single operationally interpretable\footnote{In a monthly forecasting context, the 95th percentile corresponds approximately to a one-in-20-month exceedance event, abstracting from serial dependence.} preparedness threshold, whereas the upper-tail quantile score summarizes predictive accuracy across a range of high quantiles.

We report forecast gains relative to a Gaussian no-factor Poisson random-walk benchmark. This benchmark preserves the count nature of the data and excludes the two extensions whose forecasting value we mainly aim to evaluate, namely latent common factors and flexible innovation dynamics. The reported gains therefore measure the incremental value of these extensions relative to a simple state-space benchmark. For the log predictive score, gains are additive differences relative to the benchmark, so positive values indicate improvement. For the CRPS, the 95th-percentile pinball loss, and the upper-tail quantile score, gains are percentage reductions in the average score.

\afterpage{%
\begin{landscape}
\begin{table}[p]
\centering
\caption{Country-specific best-performing one-month-ahead specifications and forecast gains relative to the Gaussian no-factor benchmark}
\label{tab:country_best_model_gains}
\centering
\resizebox{\ifdim\width>\linewidth\linewidth\else\width\fi}{!}{
\fontsize{7}{9}\selectfont
\begin{threeparttable}
\begin{tabular}[t]{lcccccccccccccccc}
\toprule
\multicolumn{1}{c}{ } & \multicolumn{4}{c}{CRPS} & \multicolumn{4}{c}{LPS} & \multicolumn{4}{c}{$q_{0.95}$ PB} & \multicolumn{4}{c}{UTQS} \\
\cmidrule(l{3pt}r{3pt}){2-5} \cmidrule(l{3pt}r{3pt}){6-9} \cmidrule(l{3pt}r{3pt}){10-13} \cmidrule(l{3pt}r{3pt}){14-17}
Country & Idio. & Factor & $Q$ & Gain & Idio. & Factor & $Q$ & Gain & Idio. & Factor & $Q$ & Gain & Idio. & Factor & $Q$ & Gain\\
\midrule
AT & SV & $t$ & 5 & +2.5\% & SV & SV & 3 & +0.130 & G & SV & 2 & +10.2\% & G & SV & 2 & +7.3\%\\
BE & SV & $t$ & 5 & +9.2\% & SV & \textemdash{} & 0 & +0.225 & SV & $t$ & 5 & +32.1\% & SV & $t$ & 5 & +31.1\%\\
BG & SV & SV & 4 & +4.8\% & SV & SV & 4 & +0.039 & SV & SV & 4 & +10.8\% & SV & SV & 4 & +10.1\%\\
CY & SV & $t$ & 4 & +5.7\% & SV & SV & 3 & +0.302 & SV & $t$ & 4 & +18.8\% & SV & SV & 5 & +18.7\%\\
CZ & SV & SV & 5 & +5.4\% & SV & \textemdash{} & 0 & +0.104 & G & SV & 3 & +18.9\% & G & SV & 3 & +18.5\%\\
\addlinespace
DE & SV & $t$ & 3 & +6.4\% & SV & G & 5 & +0.074 & SV & $t$ & 5 & +20.4\% & SV & $t$ & 5 & +18.6\%\\
DK & G & SV & 3 & +6.7\% & SV & SV & 5 & +0.581 & G & $t$ & 6 & +5.9\% & G & $t$ & 6 & +5.3\%\\
EE & SV & SV & 2 & +12.7\% & SV & $t$ & 2 & +0.168 & SV & $t$ & 5 & +23.2\% & SV & $t$ & 4 & +21.2\%\\
EL & SV & SV & 5 & +19.2\% & SV & G & 5 & +1.338 & SV & $t$ & 4 & +49.6\% & G & $t$ & 4 & +41.5\%\\
ES & SV & SV & 5 & +28.5\% & SV & G & 4 & +1.325 & SV & $t$ & 5 & +59.1\% & SV & $t$ & 5 & +56.9\%\\
\addlinespace
FI & SV & SV & 2 & +5.5\% & SV & SV & 5 & +0.345 & G & $t$ & 5 & +2.5\% & $t$ & \textemdash{} & 0 & +2.7\%\\
FR & SV & SV & 5 & +35.0\% & SV & $t$ & 2 & +1.414 & SV & $t$ & 2 & +62.1\% & SV & SV & 5 & +61.6\%\\
HR & SV & SV & 5 & +4.4\% & SV & SV & 3 & +0.052 & G & SV & 4 & +12.3\% & SV & SV & 5 & +11.0\%\\
HU & $t$ & \textemdash{} & 0 & +11.6\% & SV & SV & 4 & +0.129 & SV & G & 5 & +40.2\% & G & $t$ & 5 & +23.9\%\\
IE & SV & SV & 5 & +8.6\% & SV & SV & 5 & +0.157 & SV & $t$ & 5 & +22.4\% & SV & SV & 5 & +21.6\%\\
\addlinespace
IT & SV & $t$ & 2 & +10.2\% & SV & G & 2 & +0.364 & SV & SV & 5 & +38.9\% & SV & SV & 5 & +35.8\%\\
LT & SV & SV & 5 & +2.4\% & SV & $t$ & 6 & +0.036 & SV & G & 5 & +12.3\% & $t$ & G & 4 & +3.8\%\\
LU & SV & $t$ & 5 & +7.8\% & SV & SV & 5 & +0.160 & SV & $t$ & 5 & +27.6\% & SV & $t$ & 5 & +24.4\%\\
LV & SV & G & 6 & +2.7\% & SV & SV & 3 & +0.072 & $t$ & $t$ & 4 & +4.7\% & $t$ & $t$ & 4 & +4.9\%\\
MT & SV & G & 6 & +16.1\% & SV & SV & 3 & +0.203 & SV & $t$ & 3 & +50.2\% & SV & $t$ & 3 & +47.4\%\\
\addlinespace
NL & SV & SV & 5 & +9.3\% & SV & SV & 5 & +0.269 & SV & SV & 5 & +28.1\% & SV & SV & 5 & +25.5\%\\
PL & SV & G & 2 & +7.2\% & SV & SV & 5 & +0.185 & SV & SV & 5 & +22.5\% & SV & SV & 5 & +20.3\%\\
PT & SV & $t$ & 2 & +9.2\% & SV & SV & 6 & +0.274 & SV & SV & 5 & +24.9\% & SV & SV & 4 & +17.4\%\\
RO & SV & SV & 4 & +5.5\% & SV & SV & 3 & +0.087 & SV & SV & 4 & +6.0\% & SV & SV & 4 & +7.6\%\\
SE & SV & $t$ & 5 & +17.3\% & SV & $t$ & 5 & +0.374 & SV & G & 5 & +31.3\% & SV & G & 5 & +30.2\%\\
\addlinespace
SI & SV & $t$ & 6 & +22.3\% & SV & $t$ & 6 & +0.385 & SV & $t$ & 6 & +42.1\% & SV & $t$ & 6 & +40.6\%\\
SK & SV & SV & 2 & +2.8\% & G & SV & 6 & +0.002 & $t$ & G & 5 & +6.9\% & $t$ & G & 4 & +4.6\%\\
\bottomrule
\end{tabular}
\begin{tablenotes}
\item \textit{Notes: } For each country and scoring rule, entries report the best-performing one-month-ahead specification and its gain relative to the Gaussian no-factor benchmark. Idio. denotes the idiosyncratic innovation specification, Factor denotes the common-factor innovation specification, and \(Q\) denotes the number of latent factors. G = Gaussian, \(t\) = Student-\(t\), and SV = Gaussian stochastic volatility. For LPS, gains are additive differences relative to the benchmark; for CRPS, the 95th-percentile pinball loss, and UTQS, gains are percentage reductions in the average score. Positive values therefore indicate improved predictive performance throughout.
\end{tablenotes}
\end{threeparttable}}
\end{table}
\end{landscape}
}
\subsection{Country-specific one-month-ahead predictive performance}

Table~\ref{tab:country_best_model_gains} reports, for each country and scoring rule, the best-performing one-month-ahead specification and its gain relative to the Gaussian no-factor benchmark. The benchmark is improved upon in all country--metric combinations, although the size of the improvement varies
considerably across countries and scoring rules. Averaged across the EU-27, the best-performing specification reduces the CRPS by about 10 percent relative to the benchmark. The gains are substantially larger for the two upper-tail criteria: around 25 percent for the 95th-percentile pinball loss and around 23 percent for the upper-tail quantile score. Hence, the main empirical benefit of the richer specifications lies in forecasting unusually high asylum application counts, the part of the predictive distribution most relevant for preparedness planning.

The largest improvements are found for France, Spain, Greece, Malta, and Slovenia. For these countries, reductions in upper-tail losses often lie between 40 and 60 percent relative to the Gaussian no-factor benchmark, indicating that simple independent Gaussian dynamics understate or otherwise misrepresent the
risk of unusually large inflows. By contrast, gains are more modest for countries such as Bulgaria, Finland, Latvia, Romania, and Slovakia, suggesting that the benchmark already captures much of the predictable variation in those series.

Stochastic volatility is selected frequently for the idiosyncratic component. It is chosen for 25 countries under CRPS, 26 countries under LPS, 20 countries under the 95th-percentile pinball loss, and 18 countries under the upper-tail quantile score. This pattern indicates that time-varying country-specific uncertainty is a central feature of asylum application dynamics.

\subsection{Calibration of one-month-ahead predictive distributions}

Table~\ref{tab:quantile_coverage} reports empirical quantile coverage for the country-specific CRPS-best one-month-ahead forecasts. Since asylum applications are discrete counts, one-sided quantile coverage depends on how ties between realizations and predictive quantiles are handled. This issue is particularly relevant at lower quantiles and for countries with many zero or near-zero observations. We therefore use a mid-\(p\), or tie-adjusted, coverage measure:
\[
\widehat C_i^{\,\mathrm{mid}}(\tau)
=
\frac{1}{T_i}
\sum_{t=1}^{T_i}
\left[
\mathbf{1}\{y_{i,t+1}<\widehat q^{(1)}_{i,t}(\tau)\}
+
\frac{1}{2}\mathbf{1}\{y_{i,t+1}=\widehat q^{(1)}_{i,t}(\tau)\}
\right].
\]
This measure counts observations strictly below the predictive quantile fully and ties at the predictive quantile with weight one half. Equivalently, it is the expected coverage obtained when ties are broken at random. For discrete predictive distributions, this avoids treating all mass at the predictive quantile as either covered or uncovered.

The calibration results are generally reassuring. Median coverage is close to 50 percent for most countries, and upper-tail coverage is close to the nominal levels. Averaged across countries, empirical coverage is close to nominal in the center and upper tail, with 49.0 percent at the median, 90.4 percent at the 90th percentile, 94.8 percent at the 95th percentile, and 98.7 percent at the 99th percentile. This indicates that the selected predictive distributions provide a reasonable basis for upper-tail risk assessment, which is central to the asylum forecasting application.

\begin{table}[!t]
\centering
\caption{\label{tab:quantile_coverage}Empirical quantile coverage of country-specific CRPS-best one-month-ahead forecasts}
\centering
\resizebox{\ifdim\width>\linewidth\linewidth\else\width\fi}{!}{
\fontsize{8}{10}\selectfont
\begin{threeparttable}
\begin{tabular}[t]{lcccrrrrrrr}
\toprule
\multicolumn{4}{c}{ } & \multicolumn{7}{c}{Nominal quantile level} \\
\cmidrule(l{3pt}r{3pt}){5-11}
Country & Idio. & Factor & $Q$ & $1\%$ & $5\%$ & $10\%$ & $50\%$ & $90\%$ & $95\%$ & $99\%$\\
\midrule
AT & SV & $t$ & 5 & 2.0 & 9.5 & 19.0 & 48.0 & 87.0 & 94.0 & 97.0\\
BE & SV & $t$ & 5 & 2.0 & 7.0 & 11.0 & 44.0 & 87.0 & 94.0 & 100.0\\
BG & SV & SV & 4 & 2.0 & 8.0 & 13.0 & 48.5 & 90.0 & 94.0 & 99.5\\
CY & SV & $t$ & 4 & 3.5 & 11.0 & 15.5 & 43.0 & 91.0 & 92.0 & 99.0\\
CZ & SV & SV & 5 & 3.0 & 8.0 & 11.0 & 50.5 & 85.0 & 92.0 & 98.0\\
\addlinespace
DE & SV & $t$ & 3 & 0.0 & 5.0 & 12.0 & 52.0 & 92.0 & 96.0 & 99.0\\
DK & G & SV & 3 & 2.0 & 5.0 & 6.0 & 44.5 & 97.0 & 99.0 & 99.0\\
EE & SV & SV & 2 & 4.5 & 7.0 & 10.0 & 54.5 & 92.5 & 96.0 & 97.5\\
EL & SV & SV & 5 & 3.0 & 4.0 & 8.0 & 47.0 & 90.0 & 96.0 & 99.0\\
ES & SV & SV & 5 & 1.0 & 2.0 & 7.0 & 48.5 & 90.0 & 95.0 & 99.0\\
\addlinespace
FI & SV & SV & 2 & 1.0 & 2.0 & 11.0 & 49.5 & 91.0 & 93.0 & 98.0\\
FR & SV & SV & 5 & 2.0 & 4.0 & 4.0 & 48.0 & 90.0 & 95.0 & 99.0\\
HR & SV & SV & 5 & 1.0 & 8.0 & 12.0 & 50.0 & 91.5 & 97.0 & 100.0\\
HU & $t$ & \textemdash{} & 0 & 11.0 & 14.0 & 21.5 & 49.5 & 92.0 & 95.0 & 98.5\\
IE & SV & SV & 5 & 2.0 & 8.0 & 13.0 & 44.0 & 89.0 & 94.0 & 98.0\\
\addlinespace
IT & SV & $t$ & 2 & 3.0 & 5.0 & 8.0 & 47.0 & 90.0 & 97.0 & 99.0\\
LT & SV & SV & 5 & 1.0 & 6.0 & 9.5 & 52.0 & 89.0 & 91.0 & 99.0\\
LU & SV & $t$ & 5 & 1.0 & 3.0 & 5.0 & 47.5 & 92.0 & 99.0 & 99.0\\
LV & SV & G & 6 & 0.5 & 4.5 & 8.0 & 55.0 & 86.5 & 95.0 & 98.0\\
MT & SV & G & 6 & 3.0 & 6.0 & 11.0 & 49.0 & 93.0 & 98.0 & 99.0\\
\addlinespace
NL & SV & SV & 5 & 2.0 & 6.0 & 8.0 & 50.0 & 90.0 & 95.0 & 99.0\\
PL & SV & G & 2 & 2.0 & 4.0 & 9.0 & 42.5 & 92.0 & 95.0 & 97.0\\
PT & SV & $t$ & 2 & 1.0 & 3.0 & 8.0 & 48.0 & 90.5 & 92.0 & 99.0\\
RO & SV & SV & 4 & 2.0 & 3.0 & 5.0 & 54.0 & 90.0 & 93.0 & 98.0\\
SE & SV & $t$ & 5 & 1.0 & 4.0 & 8.0 & 53.0 & 95.0 & 95.0 & 100.0\\
\addlinespace
SI & SV & $t$ & 6 & 1.0 & 3.0 & 5.0 & 52.0 & 92.0 & 97.0 & 99.0\\
SK & SV & SV & 2 & 1.0 & 4.0 & 12.0 & 52.0 & 87.0 & 91.0 & 99.0\\
\bottomrule
\end{tabular}
\begin{tablenotes}
\item \textit{Notes: } Entries report empirical coverage probabilities, in percent, for the country-specific CRPS-best one-month-ahead model. For a calibrated predictive distribution, empirical coverage should be close to the nominal quantile level shown in the column header. Idio. denotes the idiosyncratic innovation specification, Factor denotes the common-factor innovation specification, and \(Q\) denotes the number of latent factors. G = Gaussian, \(t\) = Student-\(t\), and SV = Gaussian stochastic volatility.
\end{tablenotes}
\end{threeparttable}}
\end{table}

Calibration in the lower tail is also generally reasonable, although deviations from nominal coverage are larger. This is unsurprising for discrete count data, where lower predictive quantiles often coincide with mass points. Hungary provides the clearest example, reflecting the distinct dynamics induced by political decisions and policy actions under the Orb\'{a}n administration in the 2020s; see also Figures~\ref{fig:data-series} and~\ref{fig:data-series-log}.

Overall, the calibration results are encouraging. They suggest that a comparatively parsimonious state-space specification can produce empirically reliable short-run predictive distributions for asylum applications, despite the volatility, structural breaks, and policy sensitivity of the underlying social process. This is also visible in Supplementary Figure~\ref{fig:supp_forecast_fanplots}, which shows that the predictive distributions generally adapt to country-specific levels and periods of heightened uncertainty.

\subsection{Horizon dependence and multi-step forecast performance}
\label{sec:multistep-ahead}

\begin{table}[!t]
\centering
\caption{\label{tab:coverage_avg_horizon}Average empirical quantile coverage across the EU-27 by forecast horizon}
\centering
\fontsize{9}{11}\selectfont
\begin{threeparttable}
\begin{tabular}[t]{lrrrrrrr}
\toprule
\multicolumn{1}{c}{ } & \multicolumn{7}{c}{Nominal quantile level} \\
\cmidrule(l{3pt}r{3pt}){2-8}
Horizon & $1\%$ & $5\%$ & $10\%$ & $50\%$ & $90\%$ & $95\%$ & $99\%$\\
\midrule
One-month-ahead & 2.2 & 5.7 & 10.0 & 49.0 & 90.4 & 94.8 & 98.7\\
Three-month-ahead & 2.2 & 4.0 & 7.0 & 47.6 & 93.4 & 96.8 & 98.7\\
Six-month-ahead & 1.9 & 3.3 & 5.4 & 50.6 & 95.0 & 97.0 & 99.0\\
\bottomrule
\end{tabular}
\begin{tablenotes}
\item \textit{Notes: } 
\item Entries report mid-$p$ empirical coverage probabilities, in percent, averaged across the EU-27 destination countries. For each country and horizon, coverage is evaluated for the country-specific CRPS-best model. For a well-calibrated predictive distribution, average coverage should be close to the nominal quantile level in the column header.
\end{tablenotes}
\end{threeparttable}
\end{table}

The multi-step-ahead results show that the forecast gains documented for the one-month-ahead case are not merely a short-run phenomenon. Improvements persist at both the three- and six-month-ahead horizons, with the largest gains again concentrated in the upper tail of the predictive distribution. The complete country-level results for the three- and six-month-ahead forecasts are reported in Supplementary Tables~\ref{tab:supp_country_best_model_gains_h3}--\ref{tab:supp_quantile_coverage_h6}.

These findings highlight an additional form of horizon dependence. The preferred specification varies not only across countries and scoring rules, but also across forecast horizons. Hence, among our set of candidate models, there is no single best specification even for a fixed country and decision rule unless the relevant forecast horizon is fixed as well.

At the same time, the calibration results indicate that predictive distributions become more conservative at longer horizons. Table~\ref{tab:coverage_avg_horizon} shows that average one-month-ahead coverage is close to nominal throughout the upper tail (about 91, 95, and 98.5 percent at the 90th, 95th, and 99th percentiles), whereas at the three- and six-month horizons the upper-tail quantiles become progressively over-covered (rising to roughly 95, 97, and 99 percent six months ahead). This suggests that random-walk predictive variances may accumulate too quickly for medium-horizon forecasts. Future extensions could therefore combine the present uncertainty specification with alternative conditional mean dynamics, including covariate-driven, mean-reverting, or logistic processes, as in \citet{carammia2022forecasting} or \citet{susmann2025bayesian}.

\section{Discussion and Outlook}
\label{sec:discussion}

This paper provides, to the best of our knowledge, the first systematic evaluation of distributional accuracy metrics in short-run asylum forecasting, allowing model comparisons to be tied directly to policy-relevant loss functions. We find that flexible dynamic multivariate count models improve distributional forecasts of EU asylum applications, with the largest gains arising in the upper tail of the predictive distribution. This is the region most relevant for preparedness planning, since capacity constraints and policy costs are highly nonlinear in the size of inflows. Our empirical results carry three core findings.

First, there is no universally dominant model or factor dimension. The empirical ranking of models changes across countries, scoring rules, and forecast horizons because the scoring rules encode different forecasting objectives and the forecast horizon determines the relevant planning problem. The ``best'' model therefore depends on whether the target is overall distributional accuracy, local density at the realized value, or upper-tail capacity planning. For the European Commission or its agencies, such as the EU Asylum Agency, this implies that a centralized one-size-fits-all model selection strategy for the EU-27 can be suboptimal and may be misaligned with national capacity-planning objectives if it ignores country-specific loss functions. This is especially important in the context of the \textit{Preparedness and Crisis Blueprint} \citep{blueprint}, where optimal responses need to be tailored to the circumstances of individual countries.

Second, modeling dynamics \textit{jointly} across countries matters for distributional and tail forecasting. For CRPS, the 95th-percentile pinball loss, and the upper-tail quantile score, the selected one-month-ahead specifications include at least one latent common factor in 26, 27, and 26 of the 27 countries, respectively. This is consistent with the view that asylum flows form an interconnected migration system, in which EU-wide, origin-region, and destination-substitution dynamics can affect several countries simultaneously. For national agencies, this implies that an EU-wide joint model can be relevant even when the forecasting target is purely national. By embedding each country in the wider EU system, the joint EU-27 model allows national forecasts to borrow strength during shared shocks.

Third, the random-walk log-intensity structure appears to be a useful description of short-run national asylum application dynamics, based on the generally good one-step calibration of the selected models -- although calibration becomes more conservative at longer horizons. At least over short horizons, the results suggest that parsimonious state-space specifications are competitive forecasting tools without relying on extensive external covariate sets.

Three methodological extensions appear particularly relevant. First, future work on forecasting such systems of migration flows could make model selection more flexible by allowing the idiosyncratic specification to vary by destination country, rather than imposing it uniformly across the EU-27. This may better accommodate heterogeneity in destination-specific dynamics, although at the cost of a substantially enlarged model evaluation and selection problem. Second, future work could use dynamic model averaging to combine specifications over time instead of selecting a single model for each country and objective. This would be especially useful if the relevant forecasting environment changes across periods, for example between calm periods, regional crises, and global mobility disruptions. Third, more complex conditional (log) mean models may improve forecast accuracy at longer horizons, where the random-walk specification appears to generate increasingly conservative predictive distributions.

From an operational perspective, a natural next step is to move from methodological development to implementation in close collaboration with decision-makers and relevant agencies. This would require translating probabilistic forecasts into decision-relevant quantities that reflect the practical trade-offs involved in contingency planning.\footnote{An example is the Netherlands, where migration forecasts are used in budget planning. The Dutch budget for asylum reception has to match a bandwidth between the 30th and 70th percentiles of a forecasted cumulative probability distribution \citep{oecd2026}.} A useful starting point would be the explicit specification of loss functions that capture the asymmetric costs of under- and over-preparation, such as shortages in reception capacity versus temporarily idle resources \citep{bijak2010,zens2026dynamic}. These loss functions could be elicited jointly with policy users and migration practitioners. Since different users and agencies may face different trade-offs, the resulting decision rules may correspond to different quantiles of the predictive distribution. Such exercises could be conducted under alternative cost assumptions and compared across countries to identify where distributional forecasting offers the greatest operational value. More broadly, model-based predictive distributions could support transparent trigger mechanisms for EU-wide emergency funding allocation or temporary reception-capacity expansion. We therefore view the proposed model and forecasting experiments as a statistical foundation for more operational, agency-specific decision tools.

\bibliographystyle{elsarticle-harv}
\bibliography{asylum}

\clearpage

\appendix

\setcounter{section}{0}
\renewcommand{\thesection}{Supplement S\arabic{section}}
\setcounter{subsection}{0}
\renewcommand{\thesubsection}{\thesection.\arabic{subsection}}

\setcounter{table}{0}
\renewcommand{\thetable}{S\arabic{table}}

\setcounter{figure}{0}
\renewcommand{\thefigure}{S\arabic{figure}}

\begin{center}
    \textbf{SUPPLEMENTARY MATERIAL}

\end{center}

\section{Posterior Simulation Details}
\label{app:mcmc}

We simulate from the joint posterior distribution of the model parameters, latent
log-intensities, and latent future states using a Metropolis-within-Gibbs
algorithm. For a training sample of length \(T\) and forecast horizon \(H\), the
sampler works with the augmented latent path
\[
\boldsymbol{z}_{i,0:S}
=
(z_{i0},z_{i1},\ldots,z_{iT},z_{i,T+1},\ldots,z_{i,T+H})',
\qquad
S=T+H .
\]
Thus, for each country \(i\), the latent state vector contains an initial state
\(z_{i0}\), the \(T\) in-sample log-intensities, and \(H\) appended forecast
log-intensities. The number of latent innovations used in the variance, loading,
and factor updates is therefore
\[
S=T+H,
\]
corresponding to
\[
\Delta z_{is}=z_{is}-z_{i,s-1},\qquad s=1,\ldots,S.
\]
The Poisson likelihood is attached only to observed in-sample counts,
\(s=1,\ldots,T\). The initial state \(z_{i0}\), the appended forecast states
\(z_{i,T+1},\ldots,z_{i,T+H}\), and any genuinely missing observations, such as
the pre-2013 Croatian observations, carry no likelihood contribution and are
sampled as latent quantities.

One sweep of the sampler cycles through the following blocks.

\paragraph{(1) Latent log-intensities \(\boldsymbol{z}\)}
Conditional on the loadings, factor innovations, and idiosyncratic variances,
the state equation can be written as
\[
\Delta z_{is}=m_{is}+u_{is},
\qquad
m_{is}=\sum_{q=1}^{Q}\lambda_{iq}v_{qs},
\qquad
u_{is}\mid \psi_{is}\sim \mathcal{N}(0,\psi_{is}),
\]
with precision \(w_{is}=1/\psi_{is}\). This induces, for each country, a
Gaussian random-walk Markov field on the augmented path
\(\boldsymbol{z}_{i,0:S}\). Combined with the Poisson observation equation, the
full conditional of a single state is non-Gaussian whenever an observed count is
attached to that state. We therefore update each \(z_{is}\), \(s=0,\ldots,S\), by
single-site random-walk Metropolis.

Let
\[
L_{is}(z)=
\begin{cases}
p(y_{is}\mid z), & s=1,\ldots,T \text{ and } y_{is} \text{ is observed},\\
1, & s=0,\ s>T,\text{ or } y_{is} \text{ is missing}.
\end{cases}
\]
The Gaussian prior full conditional of \(z_{is}\), conditional on its
neighbouring states, is
\[
z_{is}\mid \cdot \ \propto\
\mathcal{N}\!\left(\mu^{z}_{is},c^{z}_{is}\right),
\]
where
\[
c^{z}_{i0}=\frac{1}{w_{i1}},
\qquad
\mu^{z}_{i0}=z_{i1}-m_{i1},
\]
for the initial state,
\[
c^{z}_{is}=\frac{1}{w_{is}+w_{i,s+1}},
\qquad
\mu^{z}_{is}
=
c^{z}_{is}
\left\{
w_{is}(z_{i,s-1}+m_{is})
+
w_{i,s+1}(z_{i,s+1}-m_{i,s+1})
\right\},
\]
for interior states \(s=1,\ldots,S-1\), and
\[
c^{z}_{iS}=\frac{1}{w_{iS}},
\qquad
\mu^{z}_{iS}=z_{i,S-1}+m_{iS},
\]
for the final appended forecast state.

We propose
\[
z_{is}^{\ast}\sim \mathcal{N}(z_{is},\tau_{is})
\]
and accept the proposal with probability
\[
\alpha_{is}
=
\min\left\{
1,\,
\frac{
L_{is}(z_{is}^{\ast})
\,
\mathcal{N}(z_{is}^{\ast}\mid \mu^{z}_{is},c^{z}_{is})
}{
L_{is}(z_{is})
\,
\mathcal{N}(z_{is}\mid \mu^{z}_{is},c^{z}_{is})
}
\right\}.
\]
The proposal scale \(\tau_{is}\) is adapted during sampling using a
Robbins--Monro update targeting an acceptance rate of \(0.234\). Since the
forecast states \(z_{i,T+1},\ldots,z_{i,T+H}\) have no likelihood contribution,
their updates are driven by the latent state equation and thereby propagate the
model forward within the MCMC algorithm.

\paragraph{(2) Idiosyncratic variances \(\psi_{is}\)}
After updating the latent states, define the idiosyncratic innovations
\[
u_{is}=\Delta z_{is}-m_{is},
\qquad s=1,\ldots,S.
\]
All variance updates are based on the \(S=T+H\) latent innovations in the
augmented path.

In the homoskedastic Gaussian specification, \(\psi_{is}=\sigma_i^2\), and the
conditional posterior is conjugate:
\[
\sigma_i^2\mid\cdot
\sim
\mathcal{IG}\left(
2.5+\frac{S}{2},
\,
1.5+\frac{1}{2}\sum_{s=1}^{S}u_{is}^2
\right).
\]

In the Student-\(t\) specification, we use the scale-mixture representation
\[
u_{is}\mid \sigma_i^2,\chi_{is}
\sim
\mathcal{N}\left(0,\frac{\sigma_i^2}{\chi_{is}}\right),
\qquad
\chi_{is}\sim\mathcal{G}\left(\frac{\nu_i}{2},\frac{\nu_i}{2}\right),
\]
so that \(\psi_{is}=\sigma_i^2/\chi_{is}\). The updates are
\[
\sigma_i^2\mid\cdot
\sim
\mathcal{IG}\left(
2.5+\frac{S}{2},
\,
1.5+\frac{1}{2}\sum_{s=1}^{S}\chi_{is}u_{is}^2
\right),
\]
and
\[
\chi_{is}\mid\cdot
\sim
\mathcal{G}\left(
\frac{\nu_i+1}{2},
\,
\frac{\nu_i+u_{is}^2/\sigma_i^2}{2}
\right).
\]
The degrees-of-freedom parameter \(\nu_i\) is updated by adaptive
random-walk Metropolis on the log scale under the shifted-exponential prior
\((\nu_i-3)\sim\mathrm{Exp}(1/6)\).

In the stochastic-volatility specification, \(h_{is}=\log \psi_{is}\) follows the
AR(1) log-volatility process described above. The complete path
\((h_{i1},\ldots,h_{iS})\) is sampled using the ancillarity--sufficiency
interweaving algorithm of \citet{kastner2016dealing}, applied to the innovation
sequence \(u_{i1},\ldots,u_{iS}\). 

\paragraph{(3) Factor loadings \(\boldsymbol{\Lambda}\)}
For \(Q>0\), given the factor innovations and idiosyncratic variances, each row
\(\boldsymbol{\lambda}_i=(\lambda_{i1},\ldots,\lambda_{iQ})'\) follows a weighted
Gaussian regression of \(\Delta z_{is}\) on
\(\boldsymbol{v}_s=(v_{1s},\ldots,v_{Qs})'\), using all \(S=T+H\) augmented
innovations. Let
\[
\boldsymbol{D}_i^{-1}
=
\mathrm{diag}\left(
\frac{1}{\tau_1\lambda^{hs}_{i1}},
\ldots,
\frac{1}{\tau_Q\lambda^{hs}_{iQ}}
\right)
\]
denote the horseshoe prior precision for row \(i\). Then
\[
\boldsymbol{\lambda}_i\mid\cdot
\sim
\mathcal{N}\!\left(
\widehat{\boldsymbol{l}}_i,
\widehat{\boldsymbol{L}}_i
\right),
\]
with
\[
\widehat{\boldsymbol{L}}_i
=
\left(
\boldsymbol{D}_i^{-1}
+
\sum_{s=1}^{S}
\frac{\boldsymbol{v}_s\boldsymbol{v}_s'}{\psi_{is}}
\right)^{-1},
\]
and
\[
\widehat{\boldsymbol{l}}_i
=
\widehat{\boldsymbol{L}}_i
\sum_{s=1}^{S}
\frac{\boldsymbol{v}_s\,\Delta z_{is}}{\psi_{is}}.
\]

\paragraph{(4) Horseshoe hyperparameters}
The local scales \(\lambda^{hs}_{iq}\), the factor-specific global scales
\(\tau_q\), and their half-Cauchy auxiliary variables are updated using the
standard conditionally conjugate inverse-gamma scheme implied by the
scale-mixture representation of the half-Cauchy distribution
\citep{makalic2015simple}.

\paragraph{(5) Factor innovations \(\boldsymbol{v}_s\)}
For \(Q>0\), conditional on \(\boldsymbol{\Lambda}\),
\(\boldsymbol{\Psi}_s=\mathrm{diag}(\psi_{1s},\ldots,\psi_{Ks})\), and
\(\boldsymbol{\Omega}_s=\mathrm{diag}(\omega_{1s},\ldots,\omega_{Qs})\), the
factor innovations are conditionally independent across \(s=1,\ldots,S\). Their
Gaussian full conditional is
\[
\boldsymbol{v}_s\mid\cdot
\sim
\mathcal{N}\!\left(
\widehat{\boldsymbol{f}}_s,
\widehat{\boldsymbol{F}}_s
\right),
\]
where
\[
\widehat{\boldsymbol{F}}_s
=
\left(
\boldsymbol{\Omega}_s^{-1}
+
\boldsymbol{\Lambda}'\boldsymbol{\Psi}_s^{-1}\boldsymbol{\Lambda}
\right)^{-1},
\]
and
\[
\widehat{\boldsymbol{f}}_s
=
\widehat{\boldsymbol{F}}_s
\boldsymbol{\Lambda}'\boldsymbol{\Psi}_s^{-1}
\Delta\boldsymbol{z}_s .
\]

\paragraph{(6) Factor variances \(\omega_{qs}\)}
The factor-variance updates mirror the idiosyncratic variance block, but the
factor scale is fixed for identification. Under the homoskedastic Gaussian
factor specification,
\[
\omega_{qs}=1
\qquad
\text{for all } q \text{ and } s.
\]
Under the Student-\(t\) factor specification, we use
\[
v_{qs}\mid \xi_{qs}
\sim
\mathcal{N}\left(0,\frac{1}{\xi_{qs}}\right),
\qquad
\xi_{qs}\sim
\mathcal{G}\left(\frac{\nu_q}{2},\frac{\nu_q}{2}\right),
\]
so that \(\omega_{qs}=1/\xi_{qs}\), with
\[
\xi_{qs}\mid\cdot
\sim
\mathcal{G}\left(
\frac{\nu_q+1}{2},
\,
\frac{\nu_q+v_{qs}^2}{2}
\right).
\]
The degrees-of-freedom parameter \(\nu_q\) is updated by adaptive Metropolis on
the log scale under the shifted-exponential prior.

Under stochastic volatility, the log factor variances follow the same AR(1)
structure as the idiosyncratic log variances, but with the log-volatility mean
restricted to zero for scale identification. The path
\((\log\omega_{q1},\ldots,\log\omega_{qS})\) is sampled using the
ancillarity--sufficiency interweaving algorithm of \citet{kastner2016dealing}.

\paragraph{Prediction}
Prediction is embedded directly in the augmented-state sampler. At each retained
draw, the appended future log-intensities \(z_{i,T+h}\), \(h=1,\ldots,H\), are
already sampled from their posterior predictive distribution conditional on the
parameters, factor innovations, and volatility paths. Count forecasts are then
obtained as
\[
y^{\ast}_{i,T+h}\sim
\mathrm{Poisson}\left\{\exp(z_{i,T+h})\right\},
\qquad
h=1,\ldots,H.
\]
Thus, multi-step predictive uncertainty is propagated through the latent
random-walk evolution, the common factor innovations, and the corresponding
idiosyncratic and factor variance processes.

\section{Additional Tables and Figures}
\label{app:multistep-tables}

\begin{figure}[H]
\centering
\includegraphics[width=\linewidth]{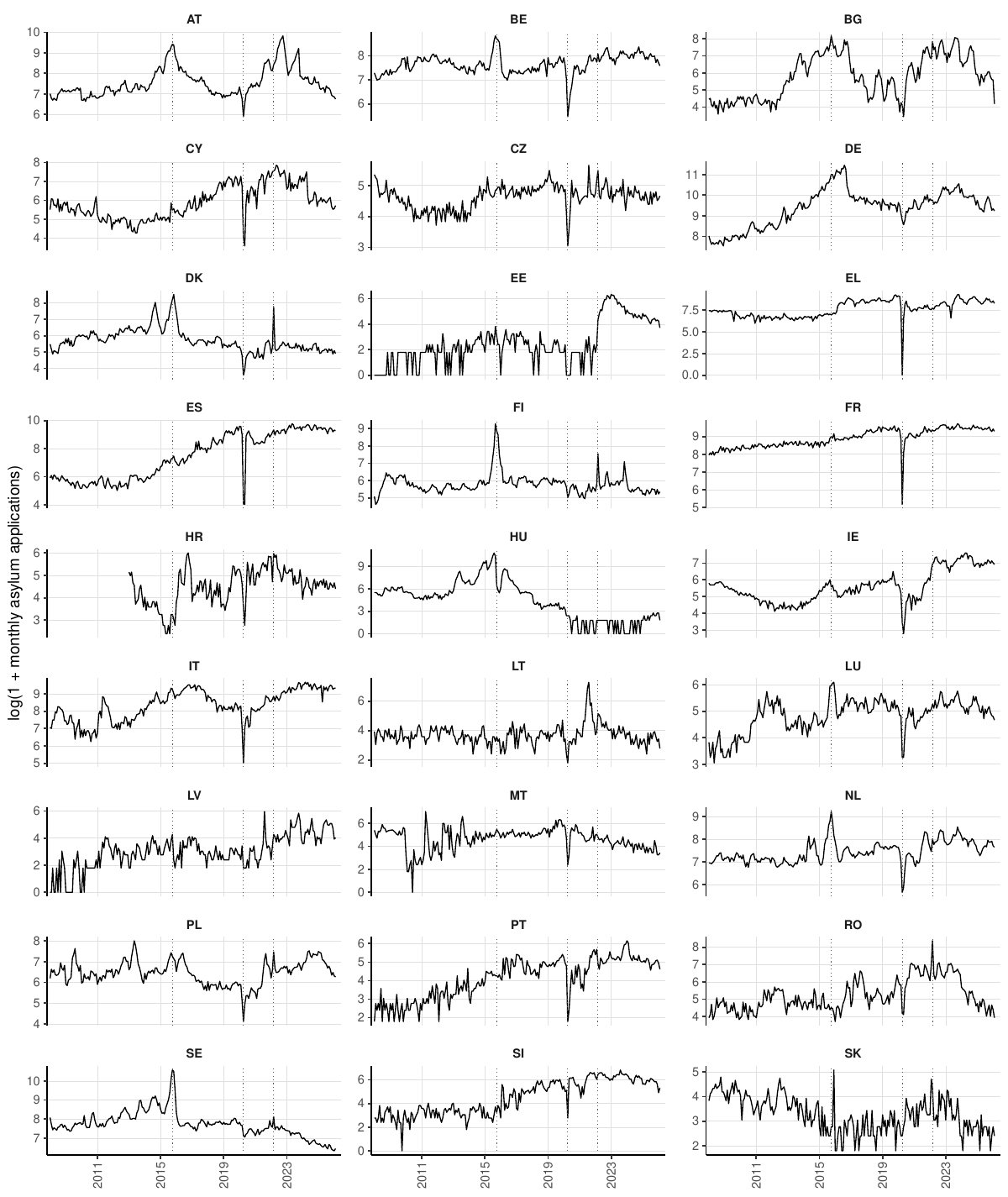}
\caption{Log-transformed monthly asylum application counts, $\log(1+y_{it})$, across the EU-27 (January 2008--February 2026). Vertical dashed lines denote major structural breaks and exogenous shocks: the peak of the Middle Eastern refugee surge (October 2015), the onset of COVID-19 mobility restrictions (April 2020), and the beginning of Ukrainian refugee inflows following the Russian invasion (March 2022). Data source: Eurostat.}
\label{fig:data-series-log}
\end{figure}

\begin{figure}[H]
\centering
\includegraphics[width=\linewidth]{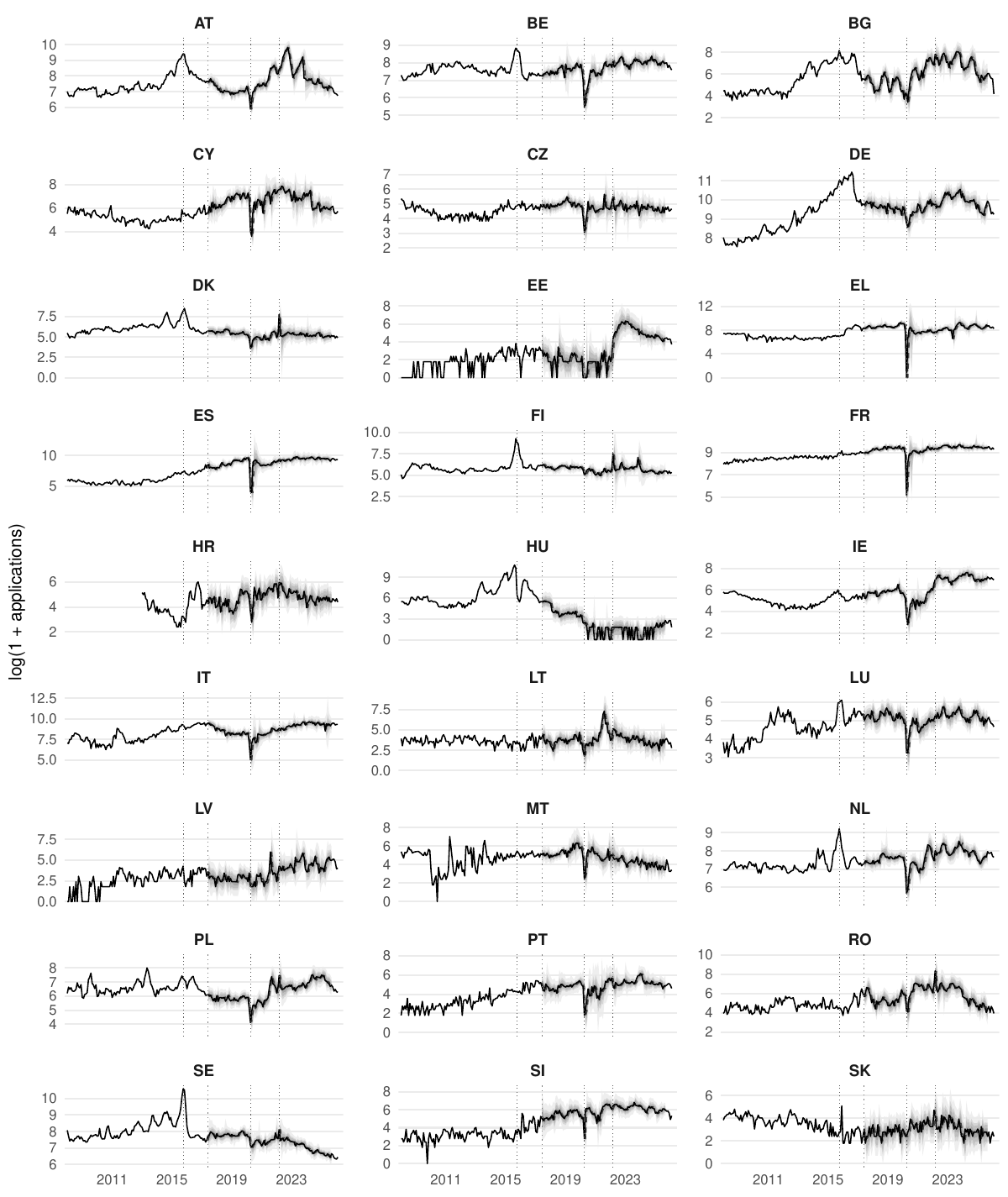}
\caption{Forecast fan plots for monthly EU-27 asylum applications. Panels show observed \(\log(1+y_{it})\) series and one-month-ahead predictive distributions. Shaded bands report posterior predictive intervals from 1--99\% to 25--75\%; point forecasts are predictive medians. Vertical dashed lines denote the start of the forecasting period as well as major structural breaks and exogenous shocks: the peak of the Middle Eastern refugee surge (October 2015), the onset of COVID-19 mobility restrictions (April 2020), and the beginning of Ukrainian refugee inflows following the Russian invasion (March 2022).}
\label{fig:supp_forecast_fanplots}
\end{figure}

\begin{table}[t]
\centering
\begin{adjustbox}{max width=\textwidth}
\begin{threeparttable}
\caption{Descriptive statistics for monthly EU-27 asylum application counts (January 2008--February 2026).}
\label{tab:sumstats}
\begin{tabular}{lrrrrrrr}
\toprule
Country & Min & Q10 & Median & Mean & Q90 & Max & SD\\
\midrule
Austria & 360 & 956 & 1,520 & 2,610.0 & 5,420 & 18,450 & 2,790.3\\
Belgium & 235 & 1,389 & 2,132 & 2,262.4 & 3,232 & 6,830 & 907.8\\
Bulgaria & 30 & 65 & 318 & 666.3 & 1,791 & 3,545 & 756.8\\
Croatia$^a$ & 10 & 30 & 95 & 115.6 & 227 & 400 & 83.9\\
Cyprus & 35 & 130 & 328 & 537.8 & 1,375 & 2,595 & 516.5\\
Czechia & 20 & 60 & 105 & 108.1 & 152 & 285 & 40.1\\
Denmark & 35 & 145 & 272 & 436.8 & 659 & 5,030 & 581.1\\
Estonia & 0 & 0 & 10 & 49.5 & 125 & 560 & 104.3\\
Finland & 100 & 204 & 310 & 462.8 & 520 & 10,815 & 951.1\\
France & 175 & 4,106 & 7,505 & 8,303.1 & 13,392 & 16,695 & 3,793.8\\
Germany & 1,880 & 2,948 & 14,110 & 17,030.6 & 31,707 & 94,360 & 15,715.1\\
Greece & 0 & 699 & 2,042 & 3,049.8 & 6,520 & 11,050 & 2,464.4\\
Hungary & 0 & 4 & 125 & 1,316.5 & 2,028 & 47,095 & 4,868.1\\
Ireland & 15 & 85 & 238 & 436.1 & 1,193 & 2,025 & 465.6\\
Italy & 150 & 1,143 & 4,512 & 5,784.5 & 12,264 & 15,740 & 4,202.0\\
Latvia & 0 & 5 & 20 & 38.9 & 88 & 390 & 54.5\\
Lithuania & 5 & 20 & 35 & 57.3 & 75 & 1,460 & 121.7\\
Luxembourg & 20 & 45 & 140 & 146.9 & 242 & 440 & 77.4\\
Malta & 0 & 25 & 125 & 147.3 & 282 & 1,130 & 130.5\\
Netherlands & 290 & 1,042 & 1,590 & 1,961.2 & 3,142 & 10,120 & 1,136.6\\
Poland & 60 & 305 & 665 & 758.9 & 1,406 & 3,010 & 449.9\\
Portugal & 5 & 10 & 80 & 97.5 & 192 & 470 & 83.8\\
Romania & 40 & 65 & 145 & 297.7 & 745 & 4,315 & 397.7\\
Slovakia & 5 & 10 & 25 & 33.4 & 70 & 160 & 25.4\\
Slovenia & 0 & 15 & 118 & 205.6 & 568 & 920 & 220.8\\
Spain & 55 & 215 & 1,850 & 4,738.6 & 12,976 & 17,600 & 5,171.8\\
Sweden & 560 & 986 & 2,218 & 3,071.5 & 4,901 & 39,060 & 4,134.0\\
\bottomrule
\end{tabular}
\begin{tablenotes}
\footnotesize
\item[a] Time series for Croatia begins in January 2013.
\end{tablenotes}
\end{threeparttable}
\end{adjustbox}
\end{table}

\clearpage
\afterpage{%

\begin{landscape}
\begin{table}[p]
\centering
\caption{Country-specific best-performing three-month-ahead specifications and forecast gains relative to the Gaussian no-factor benchmark}
\label{tab:supp_country_best_model_gains_h3}
\resizebox{\ifdim\width>\linewidth\linewidth\else\width\fi}{!}{
\fontsize{7}{9}\selectfont
\begin{threeparttable}
\begin{tabular}[t]{lcccccccccccccccc}
\toprule
\multicolumn{1}{c}{ } & \multicolumn{4}{c}{CRPS} & \multicolumn{4}{c}{LPS} & \multicolumn{4}{c}{$q_{0.95}$ PB} & \multicolumn{4}{c}{UTQS} \\
\cmidrule(l{3pt}r{3pt}){2-5} \cmidrule(l{3pt}r{3pt}){6-9} \cmidrule(l{3pt}r{3pt}){10-13} \cmidrule(l{3pt}r{3pt}){14-17}
Country & Idio. & Factor & $Q$ & Gain & Idio. & Factor & $Q$ & Gain & Idio. & Factor & $Q$ & Gain & Idio. & Factor & $Q$ & Gain\\
\midrule
AT & $t$ & $t$ & 1 & +0.9\% & SV & \textemdash{} & 0 & +0.072 & G & G & 4 & +8.1\% & G & G & 4 & +4.5\%\\
BE & SV & SV & 5 & +15.4\% & SV & SV & 5 & +0.356 & SV & SV & 5 & +41.9\% & SV & SV & 5 & +38.1\%\\
BG & G & SV & 4 & +3.9\% & $t$ & SV & 6 & +0.044 & SV & SV & 5 & +12.8\% & SV & SV & 5 & +11.9\%\\
CY & SV & $t$ & 4 & +10.1\% & SV & G & 5 & +0.337 & SV & $t$ & 4 & +26.8\% & SV & $t$ & 4 & +23.5\%\\
CZ & SV & SV & 5 & +8.3\% & SV & G & 4 & +0.196 & SV & SV & 5 & +23.8\% & SV & SV & 5 & +22.4\%\\
\addlinespace
DE & SV & SV & 5 & +8.9\% & SV & $t$ & 5 & +0.144 & SV & G & 4 & +26.5\% & SV & $t$ & 6 & +24.8\%\\
DK & G & SV & 3 & +7.6\% & SV & SV & 5 & +0.356 & G & $t$ & 6 & +10.3\% & G & $t$ & 5 & +8.8\%\\
EE & SV & $t$ & 2 & +9.5\% & SV & SV & 1 & +0.121 & SV & $t$ & 6 & +21.2\% & G & $t$ & 2 & +1.3\%\\
EL & G & $t$ & 3 & +21.1\% & SV & G & 3 & +1.573 & SV & $t$ & 4 & +53.3\% & G & $t$ & 4 & +51.4\%\\
ES & SV & $t$ & 4 & +37.5\% & SV & G & 4 & +1.788 & SV & $t$ & 5 & +65.1\% & G & $t$ & 6 & +62.6\%\\
\addlinespace
FI & G & SV & 6 & +5.9\% & SV & SV & 5 & +0.300 & G & $t$ & 6 & +8.4\% & G & $t$ & 6 & +8.5\%\\
FR & SV & $t$ & 2 & +44.0\% & SV & \textemdash{} & 0 & +1.462 & SV & $t$ & 5 & +72.9\% & SV & $t$ & 5 & +71.9\%\\
HR & SV & $t$ & 6 & +8.8\% & SV & $t$ & 4 & +0.073 & SV & SV & 5 & +15.0\% & SV & SV & 5 & +11.8\%\\
HU & G & $t$ & 6 & +13.8\% & SV & $t$ & 2 & +0.074 & SV & SV & 5 & +40.2\% & G & $t$ & 6 & +30.3\%\\
IE & SV & G & 4 & +8.5\% & SV & SV & 3 & +0.170 & SV & $t$ & 4 & +25.6\% & SV & $t$ & 4 & +23.0\%\\
\addlinespace
IT & SV & SV & 1 & +14.2\% & SV & \textemdash{} & 0 & +0.349 & SV & SV & 5 & +34.6\% & SV & $t$ & 1 & +31.7\%\\
LT & SV & $t$ & 5 & +3.4\% & SV & SV & 1 & +0.094 & $t$ & \textemdash{} & 0 & +2.1\% & $t$ & \textemdash{} & 0 & +2.7\%\\
LU & SV & $t$ & 5 & +10.3\% & SV & SV & 5 & +0.159 & SV & SV & 5 & +32.4\% & SV & SV & 5 & +32.7\%\\
LV & $t$ & G & 4 & +3.3\% & SV & SV & 4 & +0.066 & $t$ & $t$ & 3 & +6.0\% & G & $t$ & 3 & +5.0\%\\
MT & SV & $t$ & 2 & +28.8\% & SV & \textemdash{} & 0 & +0.325 & SV & $t$ & 3 & +61.7\% & SV & $t$ & 2 & +56.0\%\\
\addlinespace
NL & SV & SV & 5 & +9.2\% & SV & SV & 5 & +0.200 & SV & SV & 5 & +31.8\% & SV & SV & 5 & +31.7\%\\
PL & SV & G & 4 & +7.7\% & SV & \textemdash{} & 0 & +0.110 & SV & SV & 5 & +19.9\% & SV & SV & 5 & +17.2\%\\
PT & G & $t$ & 3 & +4.2\% & SV & SV & 5 & +0.219 & SV & $t$ & 5 & +18.0\% & G & $t$ & 3 & +11.4\%\\
RO & SV & $t$ & 2 & +4.7\% & SV & SV & 3 & +0.046 & G & $t$ & 5 & +11.3\% & G & $t$ & 6 & +10.8\%\\
SE & SV & \textemdash{} & 0 & +23.4\% & SV & $t$ & 5 & +0.416 & SV & $t$ & 3 & +44.2\% & SV & $t$ & 3 & +42.6\%\\
\addlinespace
SI & SV & $t$ & 6 & +23.4\% & SV & SV & 5 & +0.310 & SV & $t$ & 4 & +47.8\% & SV & $t$ & 6 & +41.3\%\\
SK & SV & \textemdash{} & 0 & +7.0\% & SV & $t$ & 4 & +0.121 & SV & $t$ & 6 & +17.2\% & G & $t$ & 6 & +8.0\%\\
\bottomrule
\end{tabular}
\begin{tablenotes}
\item \textit{Notes: } For each country and scoring rule, entries report the best-performing three-month-ahead specification and its gain relative to the Gaussian no-factor benchmark. Idio. denotes the idiosyncratic innovation specification, Factor denotes the common-factor innovation specification, and \(Q\) denotes the number of latent factors. G = Gaussian, \(t\) = Student-\(t\), and SV = Gaussian stochastic volatility. For LPS, gains are additive differences relative to the benchmark; for CRPS, the 95th-percentile pinball loss, and UTQS, gains are percentage reductions in the average score. Positive values therefore indicate improved predictive performance throughout.
\end{tablenotes}
\end{threeparttable}}
\end{table}
\end{landscape}
}

\begin{table}[!t]
\centering
\caption{Empirical quantile coverage of country-specific CRPS-best three-month-ahead forecasts}
\label{tab:supp_quantile_coverage_h3}
\resizebox{\ifdim\width>\linewidth\linewidth\else\width\fi}{!}{
\fontsize{8}{10}\selectfont
\begin{threeparttable}
\begin{tabular}[t]{lcccrrrrrrr}
\toprule
\multicolumn{4}{c}{ } & \multicolumn{7}{c}{Nominal quantile level} \\
\cmidrule(l{3pt}r{3pt}){5-11}
Country & Idio. & Factor & $Q$ & $1\%$ & $5\%$ & $10\%$ & $50\%$ & $90\%$ & $95\%$ & $99\%$\\
\midrule
AT & $t$ & $t$ & 1 & 5.0 & 7.0 & 9.0 & 50.0 & 91.0 & 93.0 & 99.0\\
BE & SV & SV & 5 & 2.0 & 3.0 & 7.0 & 43.0 & 91.0 & 96.0 & 100.0\\
BG & G & SV & 4 & 1.0 & 9.0 & 16.0 & 48.0 & 88.0 & 93.0 & 98.0\\
CY & SV & $t$ & 4 & 2.0 & 6.0 & 10.0 & 40.5 & 90.0 & 96.0 & 98.0\\
CZ & SV & SV & 5 & 2.5 & 3.0 & 9.0 & 50.0 & 94.0 & 96.0 & 98.0\\
\addlinespace
DE & SV & SV & 5 & 1.0 & 2.0 & 4.0 & 49.0 & 93.0 & 98.0 & 100.0\\
DK & G & SV & 3 & 3.0 & 3.0 & 5.0 & 49.5 & 96.0 & 98.0 & 99.0\\
EE & SV & $t$ & 2 & 4.5 & 5.0 & 7.5 & 54.5 & 91.5 & 95.0 & 97.5\\
EL & G & $t$ & 3 & 3.0 & 5.0 & 5.0 & 42.0 & 95.0 & 96.0 & 97.0\\
ES & SV & $t$ & 4 & 2.0 & 2.0 & 2.0 & 42.0 & 90.0 & 96.0 & 98.0\\
\addlinespace
FI & G & SV & 6 & 0.5 & 1.0 & 5.0 & 50.5 & 97.0 & 98.0 & 99.0\\
FR & SV & $t$ & 2 & 2.0 & 3.0 & 3.0 & 46.0 & 95.0 & 99.0 & 99.0\\
HR & SV & $t$ & 6 & 1.0 & 2.0 & 5.0 & 48.0 & 93.0 & 98.0 & 98.0\\
HU & G & $t$ & 6 & 11.0 & 14.0 & 18.5 & 47.5 & 95.0 & 98.0 & 100.0\\
IE & SV & G & 4 & 3.0 & 5.0 & 9.0 & 39.0 & 87.0 & 93.0 & 100.0\\
\addlinespace
IT & SV & SV & 1 & 3.0 & 6.0 & 8.0 & 52.5 & 96.0 & 99.0 & 99.0\\
LT & SV & $t$ & 5 & 0.5 & 3.0 & 5.0 & 52.5 & 93.0 & 97.0 & 98.0\\
LU & SV & $t$ & 5 & 2.0 & 2.0 & 5.5 & 46.0 & 95.0 & 97.0 & 98.0\\
LV & $t$ & G & 4 & 0.0 & 3.5 & 7.0 & 47.5 & 94.0 & 99.0 & 99.0\\
MT & SV & $t$ & 2 & 2.0 & 3.0 & 8.0 & 52.5 & 94.0 & 97.0 & 99.0\\
\addlinespace
NL & SV & SV & 5 & 3.0 & 5.0 & 8.0 & 42.0 & 92.0 & 94.0 & 98.0\\
PL & SV & G & 4 & 2.0 & 3.0 & 6.0 & 45.0 & 94.0 & 96.0 & 97.0\\
PT & G & $t$ & 3 & 2.0 & 2.5 & 6.0 & 44.5 & 97.0 & 98.0 & 99.0\\
RO & SV & $t$ & 2 & 1.0 & 5.0 & 8.0 & 53.0 & 95.0 & 96.0 & 97.0\\
SE & SV & \textemdash{} & 0 & 0.0 & 3.0 & 4.0 & 55.0 & 97.0 & 100.0 & 100.0\\
\addlinespace
SI & SV & $t$ & 6 & 1.0 & 1.0 & 4.0 & 49.0 & 94.0 & 97.0 & 100.0\\
SK & SV & \textemdash{} & 0 & 0.0 & 1.0 & 5.0 & 46.5 & 94.0 & 100.0 & 100.0\\
\bottomrule
\end{tabular}
\begin{tablenotes}
\item \textit{Notes: } Entries report empirical coverage probabilities, in percent, for the country-specific CRPS-best three-month-ahead model. For a calibrated predictive distribution, empirical coverage should be close to the nominal quantile level shown in the column header. Idio. denotes the idiosyncratic innovation specification, Factor denotes the common-factor innovation specification, and \(Q\) denotes the number of latent factors. G = Gaussian, \(t\) = Student-\(t\), and SV = Gaussian stochastic volatility.
\end{tablenotes}
\end{threeparttable}}
\end{table}


\begin{landscape}
\begin{table}[p]
\centering
\caption{Country-specific best-performing six-month-ahead specifications and forecast gains relative to the Gaussian no-factor benchmark}
\label{tab:supp_country_best_model_gains_h6}
\resizebox{\ifdim\width>\linewidth\linewidth\else\width\fi}{!}{
\fontsize{7}{9}\selectfont
\begin{threeparttable}
\begin{tabular}[t]{lcccccccccccccccc}
\toprule
\multicolumn{1}{c}{ } & \multicolumn{4}{c}{CRPS} & \multicolumn{4}{c}{LPS} & \multicolumn{4}{c}{$q_{0.95}$ PB} & \multicolumn{4}{c}{UTQS} \\
\cmidrule(l{3pt}r{3pt}){2-5} \cmidrule(l{3pt}r{3pt}){6-9} \cmidrule(l{3pt}r{3pt}){10-13} \cmidrule(l{3pt}r{3pt}){14-17}
Country & Idio. & Factor & $Q$ & Gain & Idio. & Factor & $Q$ & Gain & Idio. & Factor & $Q$ & Gain & Idio. & Factor & $Q$ & Gain\\
\midrule
AT & G & G & 4 & +0.6\% & $t$ & G & 6 & +0.020 & $t$ & G & 3 & +11.8\% & $t$ & G & 3 & +7.3\%\\
BE & SV & $t$ & 1 & +10.9\% & SV & \textemdash{} & 0 & +0.281 & SV & SV & 5 & +41.4\% & SV & SV & 5 & +37.5\%\\
BG & SV & SV & 5 & +4.1\% & G & SV & 1 & +0.026 & SV & $t$ & 5 & +12.5\% & SV & SV & 5 & +10.6\%\\
CY & G & $t$ & 5 & +11.2\% & SV & SV & 6 & +0.171 & G & $t$ & 6 & +33.4\% & G & $t$ & 6 & +32.9\%\\
CZ & SV & SV & 5 & +15.2\% & SV & SV & 5 & +0.270 & SV & SV & 5 & +31.4\% & SV & SV & 5 & +23.3\%\\
\addlinespace
DE & SV & \textemdash{} & 0 & +10.6\% & SV & $t$ & 3 & +0.154 & SV & $t$ & 6 & +36.9\% & SV & $t$ & 6 & +33.0\%\\
DK & G & SV & 3 & +7.5\% & SV & \textemdash{} & 0 & +0.338 & G & $t$ & 6 & +17.5\% & G & $t$ & 6 & +17.4\%\\
EE & G & $t$ & 2 & +3.8\% & SV & \textemdash{} & 0 & +0.109 & SV & $t$ & 6 & +18.6\% & G & $t$ & 2 & +2.0\%\\
EL & G & $t$ & 2 & +29.7\% & SV & G & 5 & +1.151 & G & $t$ & 4 & +65.6\% & G & $t$ & 4 & +65.8\%\\
ES & G & $t$ & 6 & +48.0\% & SV & SV & 6 & +1.305 & SV & $t$ & 5 & +76.6\% & G & $t$ & 5 & +75.8\%\\
\addlinespace
FI & G & $t$ & 5 & +5.6\% & SV & $t$ & 3 & +0.320 & $t$ & $t$ & 2 & +12.6\% & $t$ & $t$ & 2 & +12.3\%\\
FR & SV & $t$ & 2 & +47.0\% & SV & \textemdash{} & 0 & +1.320 & SV & $t$ & 5 & +79.1\% & SV & $t$ & 3 & +76.5\%\\
HR & SV & $t$ & 6 & +9.5\% & SV & $t$ & 4 & +0.084 & SV & $t$ & 6 & +14.0\% & SV & $t$ & 6 & +12.5\%\\
HU & G & $t$ & 6 & +15.9\% & $t$ & G & 6 & +0.038 & G & $t$ & 6 & +37.8\% & G & $t$ & 5 & +36.9\%\\
IE & SV & $t$ & 4 & +9.1\% & SV & SV & 4 & +0.147 & SV & $t$ & 6 & +22.6\% & SV & $t$ & 4 & +19.4\%\\
\addlinespace
IT & G & $t$ & 3 & +14.1\% & SV & \textemdash{} & 0 & +0.336 & SV & $t$ & 1 & +37.4\% & SV & G & 1 & +30.7\%\\
LT & G & $t$ & 1 & +1.6\% & SV & SV & 1 & +0.083 & G & $t$ & 1 & +1.4\% & G & $t$ & 1 & +1.4\%\\
LU & SV & SV & 5 & +11.7\% & SV & \textemdash{} & 0 & +0.146 & SV & SV & 5 & +34.1\% & SV & SV & 5 & +33.9\%\\
LV & G & $t$ & 5 & +5.7\% & SV & $t$ & 1 & +0.063 & G & $t$ & 5 & +8.4\% & G & $t$ & 5 & +8.9\%\\
MT & $t$ & $t$ & 4 & +12.6\% & SV & $t$ & 5 & +0.409 & SV & $t$ & 3 & +67.4\% & G & $t$ & 4 & +23.1\%\\
\addlinespace
NL & SV & $t$ & 1 & +8.6\% & SV & \textemdash{} & 0 & +0.131 & SV & SV & 6 & +26.4\% & SV & SV & 5 & +27.4\%\\
PL & SV & \textemdash{} & 0 & +7.4\% & SV & \textemdash{} & 0 & +0.109 & SV & $t$ & 4 & +16.9\% & SV & $t$ & 5 & +14.3\%\\
PT & G & $t$ & 3 & +6.3\% & SV & \textemdash{} & 0 & +0.190 & G & $t$ & 3 & +15.0\% & G & $t$ & 3 & +15.1\%\\
RO & SV & SV & 6 & +7.6\% & SV & \textemdash{} & 0 & +0.073 & G & $t$ & 6 & +19.3\% & G & $t$ & 6 & +18.5\%\\
SE & SV & \textemdash{} & 0 & +26.8\% & SV & \textemdash{} & 0 & +0.410 & SV & $t$ & 3 & +43.6\% & SV & \textemdash{} & 0 & +39.8\%\\
\addlinespace
SI & SV & $t$ & 6 & +17.1\% & SV & $t$ & 5 & +0.276 & SV & $t$ & 4 & +50.6\% & G & $t$ & 6 & +28.8\%\\
SK & G & $t$ & 6 & +4.7\% & SV & $t$ & 5 & +0.185 & SV & $t$ & 6 & +21.3\% & G & $t$ & 6 & +9.9\%\\
\bottomrule
\end{tabular}
\begin{tablenotes}
\item \textit{Notes: } For each country and scoring rule, entries report the best-performing six-month-ahead specification and its gain relative to the Gaussian no-factor benchmark. Idio. denotes the idiosyncratic innovation specification, Factor denotes the common-factor innovation specification, and \(Q\) denotes the number of latent factors. G = Gaussian, \(t\) = Student-\(t\), and SV = Gaussian stochastic volatility. For LPS, gains are additive differences relative to the benchmark; for CRPS, the 95th-percentile pinball loss, and UTQS, gains are percentage reductions in the average score. Positive values therefore indicate improved predictive performance throughout.
\end{tablenotes}
\end{threeparttable}}
\end{table}
\end{landscape}


\begin{table}[!t]
\centering
\caption{Empirical quantile coverage of country-specific CRPS-best six-month-ahead forecasts}
\label{tab:supp_quantile_coverage_h6}
\resizebox{\ifdim\width>\linewidth\linewidth\else\width\fi}{!}{
\fontsize{8}{10}\selectfont
\begin{threeparttable}
\begin{tabular}[t]{lcccrrrrrrr}
\toprule
\multicolumn{4}{c}{ } & \multicolumn{7}{c}{Nominal quantile level} \\
\cmidrule(l{3pt}r{3pt}){5-11}
Country & Idio. & Factor & $Q$ & $1\%$ & $5\%$ & $10\%$ & $50\%$ & $90\%$ & $95\%$ & $99\%$\\
\midrule
AT & G & G & 4 & 3.0 & 7.0 & 10.0 & 61.0 & 85.0 & 92.0 & 99.0\\
BE & SV & $t$ & 1 & 4.0 & 5.0 & 6.0 & 38.0 & 97.0 & 97.0 & 98.0\\
BG & SV & SV & 5 & 1.0 & 7.0 & 18.0 & 50.0 & 86.0 & 90.0 & 96.0\\
CY & G & $t$ & 5 & 2.0 & 3.0 & 5.0 & 47.0 & 96.0 & 98.0 & 98.0\\
CZ & SV & SV & 5 & 1.0 & 3.0 & 4.0 & 51.0 & 95.0 & 98.0 & 99.0\\
\addlinespace
DE & SV & \textemdash{} & 0 & 0.0 & 1.0 & 5.0 & 57.0 & 95.0 & 100.0 & 100.0\\
DK & G & SV & 3 & 1.0 & 2.0 & 3.0 & 53.5 & 96.0 & 99.0 & 99.0\\
EE & G & $t$ & 2 & 4.5 & 5.0 & 5.5 & 60.5 & 90.0 & 92.5 & 97.0\\
EL & G & $t$ & 2 & 2.0 & 4.0 & 5.0 & 45.0 & 97.0 & 97.0 & 98.0\\
ES & G & $t$ & 6 & 2.0 & 2.0 & 2.0 & 42.0 & 98.0 & 98.0 & 98.0\\
\addlinespace
FI & G & $t$ & 5 & 0.0 & 0.0 & 2.0 & 52.5 & 99.0 & 99.0 & 99.0\\
FR & SV & $t$ & 2 & 2.0 & 3.0 & 4.0 & 43.0 & 98.0 & 99.0 & 99.0\\
HR & SV & $t$ & 6 & 0.0 & 1.0 & 2.0 & 48.0 & 95.0 & 97.0 & 100.0\\
HU & G & $t$ & 6 & 11.5 & 11.5 & 18.0 & 55.0 & 97.5 & 99.0 & 100.0\\
IE & SV & $t$ & 4 & 4.0 & 6.0 & 7.0 & 41.0 & 84.0 & 90.0 & 99.0\\
\addlinespace
IT & G & $t$ & 3 & 1.0 & 2.0 & 5.0 & 43.0 & 97.0 & 99.0 & 99.0\\
LT & G & $t$ & 1 & 2.0 & 4.0 & 5.0 & 55.5 & 96.0 & 97.0 & 98.0\\
LU & SV & SV & 5 & 2.0 & 2.0 & 4.0 & 47.0 & 98.0 & 99.0 & 100.0\\
LV & G & $t$ & 5 & 0.0 & 1.0 & 1.0 & 46.5 & 98.0 & 99.0 & 99.0\\
MT & $t$ & $t$ & 4 & 0.0 & 2.0 & 2.0 & 60.5 & 98.0 & 99.0 & 100.0\\
\addlinespace
NL & SV & $t$ & 1 & 3.0 & 5.0 & 7.0 & 48.0 & 89.0 & 93.0 & 99.0\\
PL & SV & \textemdash{} & 0 & 1.0 & 3.0 & 7.0 & 45.5 & 92.0 & 94.0 & 99.0\\
PT & G & $t$ & 3 & 2.0 & 2.0 & 3.0 & 53.5 & 97.0 & 98.0 & 99.0\\
RO & SV & SV & 6 & 0.0 & 3.5 & 6.0 & 59.0 & 97.0 & 98.0 & 100.0\\
SE & SV & \textemdash{} & 0 & 0.0 & 1.0 & 5.0 & 64.0 & 99.0 & 100.0 & 100.0\\
\addlinespace
SI & SV & $t$ & 6 & 1.0 & 2.0 & 4.0 & 52.0 & 96.0 & 98.0 & 100.0\\
SK & G & $t$ & 6 & 0.0 & 0.0 & 0.0 & 48.5 & 100.0 & 100.0 & 100.0\\
\bottomrule
\end{tabular}
\begin{tablenotes}
\item \textit{Notes: } Entries report empirical coverage probabilities, in percent, for the country-specific CRPS-best six-month-ahead model. For a calibrated predictive distribution, empirical coverage should be close to the nominal quantile level shown in the column header. Idio. denotes the idiosyncratic innovation specification, Factor denotes the common-factor innovation specification, and \(Q\) denotes the number of latent factors. G = Gaussian, \(t\) = Student-\(t\), and SV = Gaussian stochastic volatility.
\end{tablenotes}
\end{threeparttable}}
\end{table}

\end{document}